\documentclass{olplainarticle}
\usepackage{subfigure}
\usepackage{caption}
\usepackage{hyperref}
\usepackage{xcolor}
\usepackage{soul}
\usepackage{mathtools}
\usepackage{amsmath, amssymb}

\title{The Theoretical Landscape of Mimetic Gravity: A Comprehensive Review}

\author[1,2]{Ola Malaeb}
\affil[1]{Department of Physics, American University of Beirut, Beirut, Lebanon}
\affil[2]{Center for Advanced Mathematical Sciences, American University of Beirut, Beirut, Lebanon}

\begin{abstract}
Mimetic gravity has emerged as a compelling extension of General Relativity (GR), originally motivated by the attempt to isolate the conformal degree of freedom of the gravitational field. By reparametrizing the physical metric in terms of an auxiliary metric and a scalar field, the theory naturally gives rise to a longitudinal degree of freedom that mimics the behavior of cold dark matter. This review provides a comprehensive survey of the theoretical landscape of mimetic gravity and its multifaceted applications to cosmology and high-energy physics. We begin by examining the original formulation and addressing the fundamental question of its equivalence to GR, highlighting how a singular disformal transformation introduces new physical degrees of freedom. We then explore minimal generalizations that lead to unified cosmological models, including mimetic matter scenarios and extensions into $f(R, \phi)$ gravity, which allow for the reconstruction of any desired expansion history. Significant attention is given to the ``limiting curvature'' hypothesis through $f(\Box \phi)$ modifications, providing a classical mechanism for resolving cosmological and black hole singularities. We critically assess the challenges facing the theory, specifically the gradient and ghost instabilities identified in cosmological perturbations, and discuss modern resolutions such as ghost-free mimetic massive gravity and covariant formulations of Ho\v{r}ava gravity. Finally, we discuss the role of the mimetic field in the early universe, specifically in the context of asymptotically free gravity and the resolution of the self-reproduction problem in inflation. 
\end{abstract}

\begin{document}

\maketitle

\section{Introduction}

More than a century after its original formulation \cite{einstein1, einstein2, einstein3, einstein4, einstein5}, General Relativity (GR) continues to stand as one of the most successful physical theories ever developed. Its predictive power has repeatedly been confirmed with extraordinary accuracy, from the deflection of starlight by massive bodies to the precession of Mercury’s perihelion, and more recently through the direct observation of gravitational waves. Yet despite this impressive record, a broad landscape of alternative or extended theories has been proposed. The motivation for these efforts arises from several conceptual and empirical challenges that suggest GR, while effective across many regimes, may not be the ultimate description of gravity. Among the most profound issues is the prediction that singularities inevitably form from well-behaved initial conditions. GR forecasts the existence of curvature singularities—such as the Big Bang at the origin of cosmic expansion and the central singularities inside black holes—where physical quantities diverge, and the classical theory loses predictive power. These extreme regimes are precisely where quantum gravitational effects are expected to dominate, highlighting the incompleteness of GR at very high energies. Indeed, at scales approaching the Planck regime, GR ceases to be reliable, reinforcing the expectation that a quantum theory of gravity is needed to describe the early universe and other high-curvature phenomena.\\

\noindent Importantly, the existence of singularities is not simply a consequence of idealized symmetric solutions. The singularity theorems developed by Penrose and Hawking \cite{penrose, hawking} proved that such divergences are generic outcomes of GR under broad and physically reasonable assumptions. Alongside this theoretical limitation, GR also faces challenges in accounting for observational data without invoking mysterious components such as dark matter \cite{darkmatter1}, \cite{darkmatter2}, and dark energy \cite{darkenergy}, which together constitute roughly 95$\%$ of the universe's total energy content \cite{planck}. This requires new ingredients beyond the Einstein–Hilbert action. These shortcomings have motivated numerous attempts to construct modified or extended theories of gravity that remain valid in extreme conditions while offering explanations for cosmic acceleration \cite{Copeland:2006wr}, structure formation \cite{Ferreira:2019xrr}, and other open problems. Proposed directions range from ultraviolet completions such as string theory \cite{Polchinski:1998rr} and loop quantum gravity \cite{Ashtekar:2004eh} to effective-field-theory approaches \cite{Burgess:2003jk} and classical modifications like $f(R)$ models \cite{DeFelice:2010aj, Sotiriou:2008rp}, scalar–tensor theories \cite{Fujii:2003pa}, and higher-dimensional frameworks \cite{Maartens:2010ar}. Many of these theories aim to avoid singularities by introducing additional physical mechanisms that regulate curvature at high energies \cite{Bojowald:2005cb, Novello:2008ra}. A particularly appealing possibility is to modify GR at sub-Planckian scales while retaining the continuum structure of spacetime. By enforcing a maximum allowable curvature \cite{Markov:1982pf, asymptotic3}, one could potentially eliminate singularities within a purely classical setting, delaying or even avoiding the need to invoke a full quantum-gravitational description.\\

\noindent A major class of modifications to GR involves enriching the gravitational sector with extra dynamical fields, most commonly scalar degrees of freedom that interact with gravity in a non-trivial way. These models are known as scalar-tensor theories \cite{tst}, and can capture phenomena beyond standard GR. While fundamental scalar particles are scarce in nature, scalar fields have proven extremely powerful in cosmology, particularly in explaining large-scale dynamics such as inflation \cite{inflation1}, \cite{newtoniangauge}, and late-time cosmic acceleration. A notable early example is Brans–Dicke theory \cite{BransDicke}, which introduces a scalar field to allow the effective gravitational coupling to vary, inspired—at least in part—by Dirac’s large number hypothesis. String theory also predicts scalar partners to the graviton, such as the dilaton, providing further theoretical motivation for these frameworks \cite{string}.\\

\noindent
More sophisticated extensions incorporate higher-derivative interactions into the gravitational action \cite{stelle_renormalization}. Although such terms are often associated with Ostrogradsky instabilities \cite{Woodard:2015zca,Ostrogradsky1850}, special constructions—such as Galileon theories—circumvent these problems by ensuring that the resulting field equations remain second order \cite{Nicolis:2008in,Deffayet:2009wt}. These models were later shown to fit within the general scalar–tensor structure formulated by Horndeski \cite{Horndeski:1974wa}, providing the most comprehensive theory of this type with second-order equations of motion \cite{Deffayet:2011gz,Kobayashi:2011nu}. Additional strategies for avoiding instabilities rely on degeneracy or constraints in the Lagrangian \cite{Langlois:2015cwa}, with mimetic gravity serving as a prominent example of this latter class.\\

\noindent The central focus of this review is mimetic gravity \cite{mimetic}, a modification of GR that reexpresses the physical metric through a constrained scalar field, thereby isolating the conformal degree of freedom. This field is special in that it can replicate the properties of dark matter, making it a flexible tool for cosmological modeling. By isolating the conformal degree of freedom, mimetic matter can explain a range of cosmological phenomena, providing a unified framework to understand both dark matter and dark energy. This review is structured to guide the reader from the foundational concepts of mimetic gravity to its most recent and advanced applications. We begin in Section \ref{section2} by detailing the original motivation behind the theory: the attempt to isolate the conformal degree of freedom of the metric tensor within the framework of general relativity. In Section \ref{section3}, we address the crucial question of whether the minimal mimetic action is truly a new theory or simply a rewriting of GR. Following this, Section \ref{section4} explores the first major application of the framework, examining cosmological consequences with mimetic matter, including its potential to act as quintessence, drive inflation without a fundamental inflaton, and realize a bouncing universe. We will also cover the analysis of cosmological perturbations in these models. The subsequent section \ref{section5} is dedicated to exploring unified cosmological models within the framework of mimetic $f(R, \phi)$ gravity. Section \ref{section6} introduces a powerful class of modifications based on functions of the form $f(\Box \phi)$, which can lead to a limiting curvature and the resolution of singularities in both cosmological and black hole spacetimes. The review then shifts toward the internal consistency of the theory; Section \ref{section7} analyzes the known instabilities in mimetic perturbations, while Sections \ref{section8} and \ref{section9} discuss theoretical resolutions via ghost-free mimetic massive gravity and mimetic Ho\v{r}ava gravity, respectively. The high-energy and early-universe implications are covered in Sections \ref{section10} and \ref{section11}, focusing on asymptotically free mimetic gravity and the dynamics of mimetic inflation. Finally, we provide concluding remarks and future outlooks in Section \ref{section12}.

\section{Attempt to Isolate Scale Factor of Metric \texorpdfstring{$g_{\mu \nu}$}{g\_mu nu} in GR}
\label{section2}

In gravitational theories, particularly scalar-tensor theories, it is common to consider different conformal frames \cite{Fujii:2003pa}. One such frame is the Jordan frame, where the scalar field couples non-minimally to the Ricci scalar. A conformal transformation of the metric is defined as
\begin{equation*}
   g_{\mu\nu}(x) = e^{2 \sigma} \, \tilde{g}_{\mu\nu}(x), 
\end{equation*}
where $\sigma$ is the conformal factor and $\tilde{g}_{\mu\nu}$ is the conformally rescaled metric. Under this transformation, the Einstein-Hilbert action transforms as
\begin{equation*}
    S = \int d^4 x \ \sqrt{-g} R(g) = \int d^4 x \ e^{2 \sigma} \sqrt{-\tilde{g}} \left( R(\tilde{g}) - 6 \tilde{g}^{\mu \nu} \left( \tilde{\nabla}_{\mu} \tilde{\nabla}_{\nu} \sigma + \partial_{\mu} \sigma \partial_{\nu} \sigma \right) \right),
\end{equation*}
where $R(g)$ and $R(\tilde{g})$ are the Ricci scalars constructed from $g_{\mu\nu}$ and $\tilde{g}_{\mu\nu}$, respectively. In the Jordan frame, scalar fields can couple to gravity in a way that mimics scale transformations, though this does not necessarily imply the scale invariance of the full action \cite{Fujii:2003pa}.\\

Building on this conformal structure, the trace-free part of the Riemann curvature tensor defines the Weyl tensor $C_{\mu\nu\rho\sigma}$. In four dimensions, this tensor captures the purely conformal part of the curvature. Specifically, under the transformation $g_{\mu\nu} \to \tilde{g}_{\mu\nu} = \Omega^2(x) g_{\mu\nu}$, the Weyl tensor with one contravariant index remains invariant \cite{Wald:1984rg}
\begin{equation*}
    \tilde{C}^{\mu}{}_{\nu\rho\sigma} = C^{\mu}{}_{\nu\rho\sigma}.
\end{equation*}
Consequently, the fully covariant form transforms as $\tilde{C}_{\mu\nu\rho\sigma} = \Omega^2 C_{\mu\nu\rho\sigma}$. This invariance allows for the construction of a locally Weyl-invariant action
\begin{equation*}
 S_{\text{Weyl}} = \alpha \int d^4x \sqrt{-g} \, C_{\mu\nu\rho\sigma} C^{\mu\nu\rho\sigma}.
\end{equation*}
While mathematically elegant, this action leads to fourth-order Bach equations rather than the second-order equations of general relativity. Although all vacuum solutions to Einstein's equations satisfy Weyl gravity, the theory generally permits non-physical solutions and suffers from the ``ghost'' problem at the quantum level, the presence of negative-norm states that threaten stability and unitarity \cite{Stelle:1977re}. Thus, it is typically viewed as a mathematical model rather than a direct physical replacement for Einstein gravity.\\

In contrast to theories that exploit conformal invariance, unimodular gravity offers a framework where the conformal degree of freedom is instead eliminated by fixing the determinant of the metric, typically as \cite{Anderson:1971pn}
\begin{equation*}
    \det g_{\mu\nu} = -1. 
\end{equation*}
This constraint restricts the group of general covariance to the subgroup of volume-preserving diffeomorphisms. Because the action is varied under this determinant constraint, the resulting field equations correspond only to the trace-free part of the Einstein equations. Crucially, in this approach, the cosmological constant $\Lambda$ emerges as an integration constant rather than a fundamental Lagrangian parameter \cite{Ellis:2010uc}. This decoupling of vacuum energy from the gravitational dynamics provides a primary motivation for the theory, offering a conceptual alternative to the cosmological constant problem without altering the local predictions of standard general relativity.\\

\noindent In 2013, Chamseddine and Mukhanov proposed to write \cite{mimetic}
\begin{equation}
    g_{\mu \nu} \equiv \tilde{g}_{\mu \nu} \tilde{g}^{\alpha \beta} \partial_{\alpha} \phi \partial_{\beta} \phi,
 \label{aux}
\end{equation}
where $\phi$ is a scalar field (now called the mimetic field) and $\tilde{g}_{\mu \nu}$ is the background metric. This construction was first introduced to covariantly isolate and dynamize the gravitational conformal degree of freedom. Its primary objective was to show that dark matter could potentially be explained through geometric properties rather than the existence of unknown particles. This specific disformal transformation is special as the physical metric $g_{\mu \nu}$ is invariant under Weyl transformations of the auxiliary metric $\tilde{g}_{\mu \nu} \rightarrow e^{\sigma} \tilde{g}_{\mu \nu}$. The action is then given by
\begin{equation*}
    \int d^4x \ \sqrt{-g(\tilde{g},\phi)} \ R(\tilde{g},\phi),
\end{equation*}
where $\phi$ and $\tilde{g}_{\mu \nu}$ are treated as independent variables. If we note from equation \ref{aux} that
\begin{equation*}
    g_{\mu \nu} \equiv \tilde{g}_{\mu\nu} (S),
\end{equation*}
where
\begin{align*}
       S &= \tilde{g}^{\alpha\beta} \partial_{\alpha} \phi \partial_{\beta} \phi \\ 
       & \Rightarrow g^{\mu \nu} \equiv \tilde{g}^{\mu\nu} (\frac{1}{S}) \\
       & \Rightarrow g^{\mu \nu}\partial_{\mu} \phi \partial_{\nu} \phi = \frac{1}{S} \left( \tilde{g}^{\mu\nu} \partial_{\mu} \phi \partial_{\nu} \phi \right) = \frac{1}{S} \ S = 1.
\end{align*}
This implies a constraint,
\begin{equation}
    g^{\mu \nu} \partial_{\mu} \phi \partial_{\nu} \phi = 1;
    \label{constraint}
\end{equation}
therefore, we have $10$ $\tilde{g}_{\mu\nu}$ and one scalar field $\phi$, but there is one constraint. Hence, this leaves 10 fields. It is noteworthy to mention that, as discussed in \cite{quanta1, quanta2, quanta3}, 3D volume quantization in noncommutative geometry leads to constraint \ref{constraint}.

\section{Is the mimetic action equivalent to GR?}
\label{section3}

The mimetic action
\begin{equation*}
    \int d^4x \ \sqrt{-g(\tilde{g},\phi)} \ R(\tilde{g},\phi)
\end{equation*} 
appears to be equivalent to GR. To check, we examine the equations of motion of $\tilde{g}_{\mu\nu}$ and $\phi$. These are respectively given by \cite{mimetic}
\begin{align}
    \left( G^{\mu\nu} - T^{\mu\nu} \right) - \left( G-T \right) g^{\mu \alpha} g^{\nu \beta} \partial_{\alpha} \phi \partial_{\beta} \phi = 0,
\label{motioneq1}
\end{align} 
\begin{align}
    \frac{1}{\sqrt{-g}} \partial_{\mu} \left( \sqrt{-g} g^{\mu \nu} \partial_{\nu} \phi \left( G-T \right) \right) = \nabla_{\mu} \left( \left(G-T \right) \partial^{\mu} \phi \right) = 0,
\label{motioneq2}
\end{align}  
where $\nabla_{\mu}$ denotes the covariant derivative with respect to the physical metric $g_{\mu \nu}$. As seen from the above equations of motion, the auxiliary metric appears via the physical metric, while $\phi$ enters the equations explicitly. Taking the trace of \ref{motioneq1} will return 
\begin{equation*}
    \left( G - T \right) \left( 1 - g^{\mu \nu} \partial_{\mu} \phi \partial_{\nu} \phi \right) = 0,
\end{equation*}
which is satisfied automatically even for $G - T \neq 0$ due to the constraint \ref{constraint}. The constraint on $\phi$ is the Hamilton-Jacobi equation defining synchronous time, $\phi = t$. In addition to the two transverse degrees of freedom associated with gravitons, the gravitational field also acquires an additional longitudinal degree of freedom. This extra mode is shared between the scalar field $\phi$ and the conformal factor of the physical metric. However, the system remains constrained by conformal invariance. To understand what this extra degree of freedom describes, we rewrite the equations of motion as
\begin{equation*}
    G^{\mu\nu} = T^{\mu\nu} + \tilde{T}^{\mu\nu},
\end{equation*}
where 
\begin{align*}
    {}& \tilde{T}^{\mu\nu} = \epsilon u^{\mu} u^{\nu} \\
    & u^{\mu} \equiv g^{\mu\alpha} \partial_{\alpha} \phi \\
    & \epsilon = G - T.
\end{align*}
 we see that the equations of motion (\ref{motioneq1} and \ref{motioneq2}) hold non-trivial solutions even in the absence of matter $(T_{\mu \nu} = 0)$. The mimetic stress-energy tensor, $\tilde{T}_{\mu \nu}$, has the same form as that of a perfect fluid with vanishing pressure ($p=0$) and energy density $\epsilon$, with the gradient of the mimetic field acts as the four-velocity. The energy density does not vanish even in the absence of matter (with no matter, $\epsilon = -R$). The mimetic fluid thus represents pressureless dust, with the mimetic constraint \ref{constraint} serving as a four-velocity normalization condition
 \begin{equation*}
     u^{\mu} u_{\mu} \equiv g^{\mu\nu} \partial_{\mu} \phi \partial_{\nu} \phi = 1.
 \end{equation*}
To verify whether the mimetic fluid can indeed serve as dust, it is necessary to examine cosmological solutions. In synchronous gauge,
\begin{equation*}
    ds^2 = dt^2 - \gamma_{ij} \left(\vec{x},t\right) dx^i dx^j
\end{equation*}
e.g. for the Friedman metric $\gamma_{ij} \equiv a^2(t) \delta_{ij}$, the $\phi$ equation of motion will give
\begin{align*}
    0 {}& = \frac{1}{\sqrt{-g}} \partial_{\mu} \left( \sqrt{-g} g^{\mu \nu} \partial_{\nu} \phi \left( G-T \right) \right) \\
    & = \frac{1}{a^3} \partial_0 \left( (G-T) a^3 \right) \\
    & \Rightarrow G-T = \frac{C(x^i)}{a^3},
\end{align*}
where the hypersurfaces of constant time were taken to be the same as the hypersurfaces for constant $\phi$ ($\phi \equiv t)$ \cite{landau}. Therefore, the energy density of the mimetic stress-energy tensor decays in proportion to the scale factor as $\frac{1}{a^3}$. Recalling that, in a Friedmann universe, the energy density of a component with a constant equation of state parameter $w$ evolves according to  
\begin{equation*}
    \frac{1}{a^{3(w+1)}};
\end{equation*}
consequently, the energy density evolution of the mimetic field corresponds to $w=0$, which is the equation of state for dust. In other words, the conformal degree of freedom in gravity can replicate the behavior of dark matter at the cosmological scale, hence the term ``mimetic dark matter". Therefore, $C(x^i)$ represents the ``amount" of this mimetic dark matter. This means, in its simplest form, where we are limiting ourselves to the Einstein action of the metric $g_{\mu\nu}$, the theory generates dark matter in the form of dust. However, we now have the freedom to add more contributions by including terms dependent on $\phi$ and its derivatives. 

\subsection{Constrained System}

We now investigate the properties of gravity based on a constrained system $(g_{\mu\nu},\phi)$. As a simplification, instead of representing $g_{\mu\nu}$ as a function of $\tilde{g}_{\mu\nu}$ and $\phi$, we constrain $g_{\mu\nu}$ with the constraint
\begin{equation*}
    g_{\mu\nu} \partial_{\mu} \phi \partial_{\nu} \phi = 1 \Leftrightarrow g_{\mu\nu} = \tilde{g}_{\mu\nu} S,
\end{equation*}
where $S = \tilde{g}^{\alpha \beta} \partial_{\alpha} \phi \partial_{\beta} \phi $. The constraint will be imposed through a Lagrange multiplier \cite{golovnev} where the mimetic action will be given by
\begin{equation}
    I = \int d^4 x \sqrt{-g} \left( -\frac{1}{2} R(g_{\mu \nu}) + \lambda \left( g^{\mu\nu} \partial_{\mu} \phi \partial_{\nu} \phi - 1 \right) \right).
\label{mimaction}
\end{equation} 

\noindent At first glance, it may seem surprising that simply by rearranging parts of the metric, without adding any new components to the action, we end up with a completely new model. This was first explained in \cite{golovnev} where this phenomenon was attributed to the variation of the action occurring within a specific subset of functions. This is the situation in mimetic gravity, as rearranging the metric components can lead to a modified version of the theory because the consistency equation forces new relations to emerge. Specifically, an extra condition is imposed on the variation of the action, requiring that
\begin{equation*}
    \int_{t_{i}}^ {t_{f}} dt \ \sqrt{\Omega} = t_{f} - t_{i},
\end{equation*}
where $\Omega \equiv \tilde{g}^{\alpha \beta} \partial_{\alpha} \phi \partial_{\beta} \phi$. Assuming, for simplicity, a spatially homogeneous variation of the form $\phi(t) = t + \delta \phi(t)$ around a spatially homogeneous field $\phi = t$, implies that $\dot{\phi} = \sqrt{\Omega}$. This type of variation imposes fewer conditions for the stationarity of the action, which in turn allows for greater freedom in the dynamics. Another interpretation was proposed in \cite{barvinsky}, where mimetic gravity is identified as a conformal (Weyl-symmetric) extension of General Relativity (GR). Also, in the latter \cite{barvinsky}, it was shown that the mimetic action \ref{mimaction} remains ghost-free, provided that the energy density of the mimetic dark matter is positive. Further, the Hamiltonian formulation of mimetic gravity was formulated in \cite{malaeb2015hamiltonian}, where it was shown that the theory is equivalent to general relativity plus an extra scalar degree of freedom that behaves like pressureless dust, with its dynamics fully captured through canonical constraints and variables. \\

\noindent The proposed modified gravity with mimetic dark matter can have many advantages over GR. A modified theory of gravity based on the constrained system $(g_{\mu\nu},\phi)$ explains dark matter in the form of dust and gives contributions to the energy-momentum tensor that depend on $\phi$ and its derivatives. Despite enormous efforts to detect new particles to explain dark matter, there is no evidence so far of new particles that can explain dark matter. Furthermore, a construction of inhomogeneous dark energy within the mimetic gravity framework was presented in \cite{inhomog}.\\

\noindent Mimetic gravity has been studied extensively in different contexts \cite{rabochaya2016note, arroja2016cosmological, cognola2016covariant, momeni2014new, zhong2018thick, zhong2019gravitational, langlois2017effective, shen2019two, fernando2020mimetic, CapelaRamazanov2014, Haghani2015, hammer2015many}. In \cite{Casalino_2018}, the first robust comparison of mimetic gravity to observational data was performed. The authors investigated the viability of mimetic gravity and its higher-order extensions (HOMim), particularly the model presented in \cite{cognola2016covariant} (a low-energy limit of projectable Hořava-Lifshitz gravity), in light of the GW170817 event. The study confirmed that the original mimetic gravity is consistent with the observed gravitational wave speed. However, their Bayesian statistical analysis of the HOMim model yields highly stringent 95\% confidence level upper limits on its free parameters, implying that these parameters should be practically zero to avoid fine-tuning and naturalness issues. 

\section{Generalizing in a Minimal Way: Cosmology with Mimetic Matter}
\label{section4}

By mimicking the behavior of dark matter and dark energy, mimetic matter provides a framework for understanding the universe's evolution from its early stages to the present day. This dual role highlights the versatility of mimetic matter in explaining the various phases of cosmic evolution.\\

\noindent Within the context of modified gravity, reference \cite{fR1} (see also \cite{fR2, fR5, mg1, mg2, myrzakulov2015inflation}) provides a generalization of mimetic gravity into the $F(R)$ framework. It was shown in \cite{fR2} how mimetic $F(R)$ gravity can realize both the early-time and late-time acceleration of the universe, while \cite{fR4} demonstrated that an inflationary era consistent with current observational data can be achieved. Similarly, in \cite{nutshell}, it was shown that both mimetic $F(R)$ and mimetic $F(G)$ can successfully reproduce the observed cosmic history, including the transition from radiation to matter domination. These extensions often require rigorous foundational checks; for instance, the Hamiltonian analysis of different mimetic models was studied in \cite{takahashi2017extended, zheng2021hamiltonian, ganz2019hamiltonian}. \\

Mirzagholi and Vikman \cite{MirzagholiVikman2015} proposed a further modification by introducing higher-derivative terms that give the mimetic field an imperfect fluid behavior with a small but nonzero sound speed, potentially leading to a unified scenario for inflation, dark matter, and dark energy. These and other features were also stressed in \cite{mimeticcosm}, the latter of which will be discussed in detail in this section. Extensive cosmological and astrophysical applications of mimetic gravity have been explored in various theoretical contexts \cite{fR1, fR2, lim2010dust, saadi2016cosmological, MirzagholiVikman2015, matsumoto2015cosmological, ramazanov2015initial, babichev2017gravitational, shen2018direct, abbassi2018anisotropic, casalino2018mimicking, nashed2023key, farsi2022structure, MyrzakulovSebastiani2015}. \\

In \cite{vagnozzi}, the focus was on whether mimetic gravity can reproduce the inferred flat rotation curves of galaxies. Vagnozzi showed that by introducing a non-minimal coupling between matter and the mimetic field, mimetic gravity can generate an extra force on test particles, leading in the weak-field limit to a Modified Newtonian Dynamics (MOND)-like acceleration law. This modification allows the theory to naturally explain flat galaxy rotation curves and the Tully–Fisher relation without invoking particle dark matter. Similarly, in \cite{staticspherically}, exact spherical solutions were derived, where the authors analytically reconstructed potentials that yield polynomial modifications to the Schwarzschild geometry. With suitable choices of these corrections, the framework can account for observed rotation curves, eliminating the need for particle dark matter. Compact stars in mimetic gravitational theory have also been studied (see, for example, \cite{mimeticstars, nashed2021, nashed2023}). Specifically, Astashenok and Odintsov \cite{neutronstars} applied realistic equations of state for neutron and quark matter, finding that mimetic gravity allows for more massive and larger-radius stars compared to General Relativity. Finally, black holes in mimetic gravity were examined in \cite{Oikonomou, oikonomou1, mimeticblackholes}. In particular, Oikonomou \cite{Oikonomou} studied black holes in mimetic $F(R)$ gravity, showing that Reissner–Nordström–Anti-de Sitter black holes can exist as solutions, and subsequently analyzed their stability and properties within this modified gravity framework.\\

\noindent In this section, we will see that we can reproduce various cosmological phenomena by generalizing the above mimetic model \ref{mimaction} in a minimal way. For a first class of modifications \cite{mimeticcosm}, we add to Einstein's action the constraint on $\phi$, and $V(\phi)$, the action will then be given by
\begin{equation*}
    I = \int d^4 x \sqrt{-g} \left( -\frac{1}{2} R(g_{\mu \nu}) + \lambda \left( g^{\mu\nu} \partial_{\mu} \phi \partial_{\nu} \phi - 1 \right) - V(\phi) + \mathcal{L}_m (g_{\mu \nu}, ...) \right).
\end{equation*}
Varying with respect to $g^{\mu\nu}$ will give
\begin{align}
    {}& G_{\mu \nu} - 2 \lambda g_{\mu \nu} + g_{\mu\nu} V(\phi) = T_{\mu\nu},
    \label{Gmunu}
\end{align}
where $G_{\mu\nu}$ denotes the Einstein tensor and $T_{\mu\nu}$ represents the energy–momentum tensor of ordinary matter. Computing the trace, we get
\begin{align*}
   {}& G - 2\lambda \partial_{\mu} \phi \partial^{\mu} \phi - 4V = T \\
   & \Rightarrow \lambda = \frac{1}{2} (G - T - 4V).
\end{align*}
Substituting the Lagrange multiplier back in equation (\ref{Gmunu}), the latter becomes
\begin{align}
    G_{\mu\nu} = (G - T - 4V) \partial_{\mu} \phi \partial_{\nu} \phi + g_{\mu\nu} V(\phi) + T_{\mu\nu}.
    \label{replacement}
\end{align}
This means that the Einstein equations are replaced by equations (\ref{replacement}) and the constraint (\ref{constraint}). Upon varying the action with respect to $\phi$ and using the expression for $\lambda$, we obtain
\begin{align}
\nabla^{\mu} \left( \left(G-T-4V \right) \partial_{\mu} \phi \right) = - V'(\phi),
\label{phieq}
\end{align}
where the prime denotes derivative with respect to $\phi$. Equation (\ref{phieq}) can also be obtained by taking the covariant derivative of equation (\ref{replacement}) and using the Bianchi identity along with the conservation of the energy-momentum tensor. Equations (\ref{replacement}) represent an extra fluid with pressure $\tilde{p} = - V$ and energy density $\tilde{\epsilon} = G - T - 3V$. \\

\noindent We look first at the solutions of equations (\ref{replacement}) along with the constraint (\ref{constraint}) in a flat universe with Friedmann metric
\begin{equation*}
    ds^2 = -dt^2 + a^2(t) \left( dr^2 + r^2 d\theta^2 + r^2 \sin^2\theta \, d\phi^2 \right).
\end{equation*}
In this case, equation (\ref{phieq}), upon assuming $T_{\mu\nu} = 0$ and setting $\phi = t$, which is a general solution of the constraint equation, gives the below
\begin{align}
    {}& \frac{1}{a^3} \frac{d}{dt} \left( a^3 (\tilde{\epsilon} - V) \right) = - \frac{d{V}}{dt} \nonumber \\
    & \Rightarrow \tilde{\epsilon} = V - \frac{1}{a^3} \int dt (a^3 \dot{V}) = \frac{3}{a^2} \int a^2 V da.
\label{density}
\end{align} 
Note that in synchronous gauge, $V(\phi) = V(t)$, which appears to break diffeomorphism invariance, but it does not. Integrating the first equation in (\ref{density}) gives the second equation, which is an expression of the energy density as a function of the potential $V$. With a non-vanishing potential V, there is an additional mimetic matter contribution, analogous to the inclusion of a cosmological constant term in the Lagrangian. Now, let us look at the $0-0$ component of equation (\ref{replacement}), we get
\begin{align}
    {}& G_{00} = (G - T - 4V) + g_{00} V + 0 \\
    & \Rightarrow H^2 = \frac{1}{3} \tilde{\epsilon} = \frac{1}{a^3} \int a^2 V da
    \label{H2}
\end{align}
where $H \equiv \frac{\dot{a}}{a}$. To simplify the above integral equation, we multiply it by $a^3$ and differentiate it with respect to time, which gives us
\begin{equation}
    2\dot{H} + 3 H^2 = V(t).
\label{equationH}
\end{equation}
This can be written as a linear differential equation, after denoting $y=a^{3/2}$,
\begin{equation}
    \ddot{y} - \frac{3}{4} V(t) y = 0,
\label{DE}
\end{equation}
which is easily solvable. This second-order linear differential equation allows one to find cosmological solutions. Many cosmological models can be easily reproduced without adding new scalar fields, as we can control the behavior of the energy density by cleverly choosing the potential $V$. In the following subsections, we are going to consider several potentials. 

\subsection{Mimetic matter as quintessence} 

Here, we investigate the behavior of mimetic matter in a universe predominantly filled with other matter characterized by a constant equation of state $p = w \epsilon$, with the potential specified by
\begin{equation*}
  V(\phi) = \frac{\alpha}{\phi^2} = \frac{\alpha}{t^2}.
\end{equation*}
In this case, it can be shown that the scale factor $a(t)$ evolves with time as
\begin{equation}
    a(t) \propto t^{\frac{3}{2}(1 + w)}
\end{equation}
and the energy density of mimetic matter, assuming $\phi = t$, decays as
\begin{align*}
& \tilde{\epsilon} = -\frac{\alpha}{w t^2}.
\end{align*}
Mimetic matter follows the same state equation as the universe's dominant matter. The total energy density $\epsilon$ is given by
\begin{equation*}
    \epsilon = 3H^2 = \frac{4}{3(1 + w)^2 t^2}
\end{equation*}
where H is the Hubble parameter. The behavior of mimetic matter can be subdominant if $\frac{\alpha}{w} \ll 1$. Here mimetic matter is acting as quintessence, and for $\alpha \gg 1$, we get $\tilde p = - \tilde \epsilon$, the equation of state approaches the cosmological constant. 

\subsection{Inflation with no Inflation Field}

Here we explore the case where the inflationary solutions are constructed using the mimetic matter. By introducing a suitable potential for the mimetic scalar field, mimetic matter can act as the inflaton field, driving the inflationary phase of the universe. If one takes the scale factor $a(t) = y^{2/3}$, then the potential can be found by rewriting equation (\ref{DE}) as
\begin{equation*}
   V(\phi) = V(t) = \frac{4 \ddot{y}}{3 y}.
\end{equation*}
Inflation is then described with a graceful exit to matter dominating universe when the potential is 
\begin{equation*}
    V(\phi) = \frac{\alpha \phi^2}{\exp(\phi) + 1},
\end{equation*}
with $\alpha$ being positive. In this case, the scale factor a(t) evolves with time as
\begin{align*}
    a(t) \propto exp \left( -\sqrt{ \frac{\alpha}{12} } t^2 \right)
\end{align*}
at large negative t, while it is proportional to $t^{2/3}$  for positive t (universe dominated by dust).

\subsection{Bouncing Universe}

Mimetic matter can lead to a non-singular bouncing universe. This means that instead of a singularity, the universe undergoes a bounce, avoiding infinite density and temperature, and providing a smooth transition from contraction to expansion. This model provides a compelling alternative to traditional cosmological theories. If the potential is
\begin{align*}
    V(\phi) = \frac{4}{3} \frac{1}{(1 + \phi^2)^2},
\end{align*}
then the corresponding scale factor is   
\begin{align*}
    a = \left( \sqrt{1 + t^2} \ \left( 1 + A \ arc tan \ t\right) \right)^{2/3}    
\end{align*}
where $A$ is a constant obtained after solving the differential equation and using the freedom for normalization of the scale factor in a flat universe. If we take $A = 0$ for simplicity, the scale factor will become
\begin{equation*}
    a = (t^2 + 1)^{1/3},
\end{equation*}
and then the pressure and energy density are given respectively by
\begin{align*}
    \tilde{p} = - \frac{4}{3} \frac{1}{(1+t^2)^2}, \quad \tilde{\epsilon} = 3H^2 = \frac{4}{3} \frac{t^2}{(1+t^2)^2}.
\end{align*}
When t is large and negative, the universe behaves like a dust-dominated system with minimal pressure, causing it to contract. Initially, the energy density increases in proportion to $a^{-3}$, but during a brief period around $|t| \sim 1$, it rapidly decreases to nearly zero. This marks the point where the contraction halts, and the universe begins to expand. Following the start of expansion, the energy density quickly rises to the Planckian level over a short time interval. Afterward, the universe continues expanding in a manner typical of a dust-dominated phase. A noteworthy feature of this model is the transition in the equation of state, from one where $\tilde{\epsilon} + \tilde{p} > 0$ (normal matter) to a phantom state where $\tilde{\epsilon} + \tilde{p} < 0$. The sum of the energy density and pressure is given by
\begin{equation*}
    \tilde{\epsilon} + \tilde{p} = \frac{4}{3} \frac{t^2 - 1}{(1+t^2)^2}.
\end{equation*}
For $|t| < 1$, mimetic matter behaves like a phantom field, allowing for the possibility of a non-singular bounce. In the general scenario where $A \neq 0$, the bounce remains non-singular if $|A| < 2/\pi$. However, if $|A| \geq 2/\pi$, the universe undergoes a contraction that leads to a singularity. The value of A is determined by the relative contributions of phantom and dark matter ``components" in the mimetic matter near the point of bounce. By using this phantom mimetic matter, the singularity in a contracting universe can be avoided, even in the presence of other types of matter. In the model we examined, the bounce occurs at Planck scales, where the theory is not well-defined due to quantum gravitational effects. However, by making a slight adjustment to the potential $V$, we can reduce the bounce scale and extend the duration of the bounce beyond the Planck time. In fact, considering the potential 
\begin{equation*}
    V(\phi) = \frac{4}{3} \frac{\alpha}{(t_0^2 + t^2)^2},
\end{equation*}
will induce the bounce at scales of approximately $\alpha t_0^{-2}$ over the time interval $t_0$.

\subsection{Global Dynamical Analysis and Stability}

While the reconstruction methods introduced by Chamseddine and Mukhanov in \cite{mimeticcosm} demonstrate that mimetic gravity can realize any desired expansion history $a(t)$ through a suitable potential $V(\phi)$, a complete understanding of the theory requires investigating whether these solutions are stable attractors in the cosmic evolution. In \cite{Dutta2018}, a comprehensive dynamical systems analysis was carried out to systematically explore the cosmological implications of mimetic gravity beyond specific reconstructed solutions. To study the global behavior of the theory, the authors reformulate the field equations as an autonomous dynamical system by introducing the following dimensionless variables
\begin{equation*}
    x = \frac{\kappa^2\lambda}{3H^2}, \qquad y = \frac{\kappa\sqrt{V}}{\sqrt{3}H}, \qquad \sigma = \frac{\kappa\sqrt{\rho}}{\sqrt{3}H},
\end{equation*}
subject to the Friedmann constraint $x + y^2 + \sigma^2 = 1$. Here, $\kappa^2 = 8\pi G$ is the gravitational coupling constant and $\rho$ denotes the energy density of the standard-model matter fluid (such as radiation or dust). The evolution of the scalar potential is encapsulated through the variables,
\begin{equation*}
    z = -\frac{1}{\kappa V^{3/2}}\frac{dV}{d\phi}, \qquad \Gamma = \frac{V V_{\phi\phi}}{V_\phi^2},
\end{equation*}
allowing for a unified treatment of a broad class of potential forms. The subscript $\phi$ denotes differentiation with respect to $\phi$.\\

\noindent The resulting phase-space analysis reveals that the system naturally admits matter-dominated solutions (where $\Omega_\lambda \simeq 1$ or $\Omega_m \simeq 1$) which typically behave as saddle points. This is a crucial result, as it allows the universe to remain in a non-accelerating matter era for a sufficiently long transient period—facilitating the growth of large-scale structure—before eventually evolving toward late-time attractors. Furthermore, scalar-field dominated de Sitter solutions with $\Omega_\phi = 1$ and $w_{\rm eff} = -1$ naturally arise as stable late-time states for various potentials. \\

\noindent A particularly significant finding in \cite{Dutta2018} is the existence of accelerated scaling solutions, where the energy density parameters for dark matter and dark energy are of the same order, $\Omega_\lambda \sim \Omega_\phi = \mathcal{O}(1)$, with an effective equation of state $w_{\rm eff} < -1/3$. This provides a dynamical mechanism to alleviate the cosmic coincidence problem. Unlike conventional interacting dark-energy models, this effective interaction emerges intrinsically from the mimetic constraint and the geometry of the scalar potential.\\

\noindent The general framework is illustrated through several specific potential choices. For the inverse-square potential, $V(\phi) = \alpha \phi^{-2}$ (which implies $\Gamma = 3/2$), the system reduces to a two-dimensional phase space and yields stable, scaling accelerated attractors that successfully reproduce the sequence of radiation domination, mimetic matter domination, and late-time acceleration. Other choices, such as power-law ($V \propto \phi^n$) and exponential ($V \propto e^{\alpha\phi}$) potentials, also lead to viable cosmological histories, though they often evolve toward de Sitter states at infinity, requiring center manifold theory to verify their asymptotic stability. Additionally, upon exploring the potential $V(\phi) \propto (1 + \beta \phi^2)^{-2}$, one can demonstrate that it admits a phantom-like late-time acceleration ($w_{eff} < -1$). Overall, this dynamical analysis provides the necessary mathematical rigor to confirm that the dark-sector unification proposed in the original mimetic framework is a robust and stable feature of the theory across all cosmic epochs.

\subsection{Cosmological Perturbations and the First Class of modifications}

We now turn to the analysis of small longitudinal metric perturbations in a universe dominated by mimetic matter. The metric of the perturbed universe in the Newtonian gauge can be expressed as follows \cite{newtoniangauge}
\begin{equation*}
    ds^2 = (1 + 2 \Phi) dt^2 - (1 - 2\Phi) a^2 \delta_{ij} \  dx^i dx^j,
\end{equation*}
where $\Phi$ is the Newtonian gravitational potential. We will now consider perturbations of the scalar field. To see how the mimetic field affects perturbations of scalar modes, we start by considering
\begin{equation*}
    \phi = t + \delta \phi.
\end{equation*}
The constraint on $\phi$ (\ref{constraint}) gives $\delta \dot{\phi} = \Phi$. This implies
\begin{align*}
    {}& \delta \ddot{\phi} + H \delta \dot{\phi} + \dot{H} \delta \phi = 0 \\
    & \Rightarrow \Phi = \delta \dot{\phi} = A \left(1 - \frac{H}{a} \int a dt \right).
\end{align*}
This is precisely the general solution typically found for long-wavelength cosmological perturbations; however, in this case, this solution remains universally valid for all perturbations, independent of their wavelength. In this sense, the perturbations behave like dust with zero sound speed, even when the mimetic matter possesses nonzero pressure. Therefore, when the mimetic model is applied to the early universe, it fails to generate the primordial perturbations from quantum fluctuations necessary for forming large-scale structures. This means that different contributions should be added.\\

\noindent To address the issue discussed above, the failure of the mimetic inflationary scenario to generate the initial homogeneities required for large-scale structure formation, higher-derivative terms of the mimetic field were introduced in \cite{mimeticcosm}, modifying the action to become 
 \begin{equation*}
    I = \int d^4 x \sqrt{-g} \left( -\frac{1}{2} R(g_{\mu \nu}) + \lambda \left( g^{\mu\nu} \partial_{\mu} \phi \partial_{\nu} \phi - 1 \right) - V(\phi) + \frac{1}{2} \gamma \left( \Box \phi \right)^2 \right)
 \end{equation*}
 where $\gamma$ is a constant and $\Box = g^{\mu \nu} \nabla_{\mu} \nabla_{\nu}$. By varying the action with respect to the metric, we get the following equations of motion
\begin{equation*}
    G^\mu_{\ \nu} =  \left( V + \gamma \left( \phi_{,\alpha} \chi^{,\alpha} + \frac{1}{2} \chi^2 \right) \right) \delta^\mu_{\ \nu}
    + 2\lambda \phi_{,\nu} \phi^{,\mu}
    - \gamma \left( \phi_{,\nu} \chi^{,\mu} + \chi_{,\nu} \phi^{,\mu} \right),
\end{equation*}
\begin{equation*}
    \chi = \Box \phi.
\end{equation*}
In the Friedmann universe, the general solution of the constraint equation, eq. \ref{constraint}, is 
\begin{equation*}
    \phi = t + A,
\end{equation*}
and then $\chi$ will become
\begin{equation*}
    \chi = \Box \phi = \ddot{\phi} + 3H \dot{\phi} = 3 H \ .
\end{equation*}
The equations of motion then reduce to the following, instead of equation (\ref{equationH}),
\begin{equation*}
    2\dot{H} + 3H^2 = \frac{2}{2 - 3\gamma} V.
\end{equation*}
Upon analyzing cosmological perturbations within a Friedmann-Lemaître-Robertson-Walker (FLRW) background, the resulting equation for scalar perturbations (for details, check \cite{mimeticcosm})
\begin{equation*}
    \delta \phi_k'' + \left( c_s^2 k^2 + \frac{a''}{a} - 2\left( \frac{a'}{a} \right)^2 \right) \delta \phi_k = 0,
\end{equation*}
where the prime denotes the derivative with respect to the conformal time ($\eta = \int dt/a$), deviates from the standard form due to a gradient term arising from the mimetic modifications. This leads to a non-trivial sound speed $c_s$ for scalar modes.  By canonically quantizing short-wavelength quantum fluctuations and subsequently matching them to their long-wavelength counterparts, the amplitude of the gravitational potential $\Phi_k$ is derived. For $c_S \ll 1$, the behavior is similar to k-inflation, while for $c_S \gg 1$, the effect is opposite to what happens in k-inflation. The typical amplitude of the gravitational potential is $c_S^{1/2}$ enhanced with respect to the amplitude of the gravity waves. In this mimetic model, scalar perturbations consistently dominate over tensor perturbations. Notably, the suppression of gravitational waves arises solely from the quadratic term involving the d’Alembertian operator. As a result, unlike in k-inflation scenarios, this suppression is not expected to generate any non-Gaussianity.\\

\noindent We saw above that the introduction of higher-derivative terms effectively renders the mimetic field $\phi$ dynamical. However, the question is whether the model discussed above is ghost-free because usually the higher-derivative terms lead to the Ostrogradsky instability \cite{chen2013higher}. For instance, the $\Box \phi$ term in the action introduces two modes: one normal and the other a ghost, characterized by a kinetic term with the wrong sign. The mimetic constraint, though, eliminates one dynamical degree of freedom. What about the other? In reference \cite{Ijjas}, it was demonstrated through an action-based analysis that, despite avoiding ghost instabilities in a suitable parameter range, the model still suffers from a gradient instability in its scalar perturbations. This type of instability implies an imaginary sound speed, leading to rapid growth of perturbations across all scales and rendering linear perturbation theory invalid. Similar instabilities have also been noted \cite{Ramazanov, BenAchour:2016, Chaichian:2014}. Furthermore, it was shown in \cite{Firouzjahi} that these issues cannot be resolved merely by replacing the $\gamma (\Box \phi)^2$ term with a general function $f(\Box \phi)$ in the action. To develop a viable early-universe model based on mimetic matter, it is essential to eliminate both ghost and gradient instabilities. This challenge was also addressed in \cite{Zheng}, where it was proposed that including direct couplings between high-derivative terms and curvature may help overcome these instabilities. However, subsequent work in \cite{zheng2021hamiltonian} indicated that the theory possesses an additional scalar degree of freedom. While this mode remains absent in purely homogeneous settings, it can trigger instabilities when considering perturbations against a Minkowski background with a non-homogeneous field \cite{Ganz:viable}.

\section{Unified Cosmology in Mimetic \texorpdfstring{$f(R, \phi)$}{f(R, phi)} Gravity}
\label{section5}

While the original formulation of mimetic gravity was motivated by the dark matter problem within general relativity, much work has explored its compatibility with extended theories of gravity to provide a complete description of the cosmic history. A notable example is the work done in \cite{myrzakulov2015inflation}, where it was demonstrated that inflation can be consistently realized within a broad class of $f(R,\phi)$ theories and subsequently embedded into mimetic gravity. This approach aims to unify early-time inflation, late-time dark energy, and dark matter within a single covariant action.\\

\noindent In this extension, the authors utilize the Lagrange multiplier formulation. The action describes a gravity sector non-minimally coupled to a scalar inflaton field $\phi$, while the mimetic degree of freedom is encoded via a second scalar field $\varphi$ constrained to satisfy the condition $g^{\mu\nu}\partial_\mu\varphi\partial_\nu\varphi = -1$. The action is given by \cite{myrzakulov2015inflation}
\begin{equation}
\label{eq:mimetic_fRphi_action}
S = \int_{\mathcal{M}} d^4x \sqrt{-g} \left[ \frac{f(R, \phi)}{2} - \frac{\omega(\phi)}{2}g^{\mu\nu}\partial_\mu\phi\partial_\nu\phi - V(\phi) + \lambda(g^{\mu\nu}\partial_\mu\varphi\partial_\nu\varphi + 1) \right],
\end{equation}
where $\lambda$ is the Lagrange multiplier, $V(\phi)$ is the inflaton potential, and $\omega(\phi)$ characterizes the kinetic coupling of the inflaton. Note that $\varphi$ represents the mimetic field (identified with cosmological time), while $\phi$ is the dynamical field driving inflation, and it couples nontrivially to curvature.\\

\noindent Varying the action with respect to the physical metric $g_{\mu\nu}$ yields the modified Einstein field equations
\begin{equation}
G_{\mu\nu} = \frac{1}{F(R, \phi)} \left( T^\phi_{\mu\nu} + T^{\text{MG}}_{\mu\nu} + \tilde{T}_{\mu\nu} \right),
\end{equation}
where $F(R, \phi) \equiv \partial f / \partial R$. The source terms include the standard stress-energy tensor for the inflaton ($T^\phi_{\mu\nu}$) and a geometric contribution from the modified gravity sector
\begin{equation}
T^{\text{MG}}_{\mu\nu} = \frac{g_{\mu\nu}}{2}\left(f(R, \phi) - R F(R, \phi)\right) + \nabla_\mu\nabla_\nu F(R, \phi) - g_{\mu\nu}\Box F(R, \phi).
\end{equation}
The distinct feature of the theory is the mimetic stress-energy tensor $\tilde{T}_{\mu\nu}$, which arises from the constraint term and takes the form
\begin{equation}
\tilde{T}_{\mu\nu} = -(F G - T^\phi - T^{\text{MG}}) \partial_\mu\varphi\partial_\nu\varphi,
\end{equation}
where $G$, $T^\phi$, and $T^{\text{MG}}$ denote the traces of the Einstein tensor and the respective stress-energy tensors.\\

\noindent 
Myrzakulov \emph{et al.} applied this framework to a scenario in which the cosmological constant is promoted to a dynamical quantity through its coupling to a scalar field. In particular, they considered models in which the curvature scale that controls the effective cosmological constant is replaced according to $R_0^{-1}\rightarrow -b\kappa^{3}\phi$, leading to the exponential gravity form
\begin{equation}
f(R,\phi)=\frac{R-2\Lambda\!\left(1-e^{b\phi\kappa^{3}R}\right)}{\kappa^{2}} .
\end{equation}
In this construction, the scalar field $\phi$ controls a running curvature scale that determines the magnitude of the effective cosmological constant. In the high-curvature regime the theory reduces to $f(R,\phi)\simeq R-2\Lambda_{\rm eff}(\phi)$ and supports a quasi-de Sitter inflationary phase with $H^{2}\simeq \Lambda_{\rm eff}/3$. As $\phi$ evolves, the effective cosmological constant is dynamically suppressed, and the theory is driven toward the Einstein-Hilbert limit $f(R,\phi)\to R$, providing a graceful exit from inflation. The authors analyzed both exponential 
\begin{equation}
f(R, \phi) = \frac{R - 2\Lambda(1 - e^{b\phi\kappa^3 R})}{\kappa^2},
\end{equation}
and Hu-Sawicki-type realizations and showed, using the generalized slow-roll formalism for $f(R,\phi)$ gravity, that the models predict a red-tilted scalar spectrum with
\begin{equation}
1-n_{s}\simeq \frac{2}{N}, \qquad r=\mathcal{O}(N^{-4}),
\end{equation}
so that the tensor-to-scalar ratio is strongly suppressed, making these mimetic inflationary scenarios consistent with current observational constraints.\\

\noindent Despite its conceptual appeal in unifying inflation and dark matter, this mimetic $f(R, \phi)$ framework faces several theoretical challenges. A primary issue is the exponential dilution of any pre-existing mimetic dark matter during the inflationary phase. To address this, the seeding mechanism originally proposed by Chamseddine and Mukhanov \cite{mimeticcosm} for standard mimetic gravity—wherein the mimetic density is continually generated via the scalar potential—is essentially required to ensure a realistic post-inflationary abundance in $f(R, \phi)$ realizations. Furthermore, the model requires a robust perturbative analysis to be considered a viable candidate for structure formation. In the minimal formulation, the sound speed of scalar perturbations is identically zero ($c_s = 0$), leading to the formation of caustic singularities and the inability to generate primordial fluctuations. To resolve this within the modified gravity context, higher-derivative extensions of mimetic $f(R)$ gravity, often incorporating $(\Box \varphi)^2$ terms, were investigated by authors such as Nojiri and Odintsov \cite{fR1} and Arroja \emph{et al.} \cite{Arroja2015}. These works demonstrated that a non-zero sound speed can be achieved, allowing for a healthy growth of perturbations. \\

However, more recent systematic studies have shown that $f(R)$ and $f(R, \phi)$ extensions of mimetic gravity are not immune to the instabilities that plague the original theory. Specifically, the perturbative analysis by Ijjas and Steinhardt \cite{Ijjas} and Ramazanov \emph{et al.} \cite{Ramazanov} highlighted that many such extensions suffer from ghost or gradient instabilities (imaginary sound speed) in the scalar sector. These findings have forced a move toward even more complex generalizations, such as the non-minimal derivative couplings or the DHOST-mimetic theories investigated in \cite{Casalino_2018}, to ensure that the unified expansion history remains consistent with both the mathematical stability requirements and the observational data from the CMB.

\section{Second Class of Modifications \texorpdfstring{$f(\Box \phi)$}{f(Box phi)}: Theory With Limiting Curvature} 
\label{section6}

General relativity not only predicts the existence of black holes but also implies a singularity that is hidden by the event horizon. At this singularity, spacetime curvature becomes infinite, and space-times cannot be geodically complete \cite{penrose}. According to GR, an infalling observer would cross the event horizon in finite proper time and continue falling inevitably toward the singularity. Before reaching it, however, the observer would be stretched and torn apart by infinite curvatures \cite{blackholeintro}. Hawking and Penrose \cite{penrose} proved that the formation of singularities is an inevitable feature of general relativity, not just an artifact of symmetry. However, it is well known that Penrose's theorem can be violated because the conditions are not always valid. For example, a scalar field condensate or a cosmological constant violates some of the conditions, and hence, allows spacetime to avoid a singularity and remain geodesically complete.\\

\noindent In this section, we will study the attempt to resolve singularities entirely at the classical level, as done in \cite{cosmosingu}, by introducing the idea of a limiting curvature \cite{limiting1}, \cite{limiting2}, \cite{limiting1}, \cite{limiting2}, \cite{limiting3}, \cite{limiting4}, assuming that Einstein’s equations are modified at curvatures well below the Planck scale. Since General Relativity has only been tested in regimes of relatively low curvature, such modifications cannot be forbiden. If the limiting curvature is sufficiently low, quantum effects—such as particle production and vacuum polarization—may be safely neglected. In this way, the theory remains well-behaved and reliable at extremely high curvatures while still being consistent with experimental results at low curvature. Research in this direction has seen some development; for instance, references \cite{limiting3} and \cite{limiting4} proposed extending Einstein's equations by incorporating intricate combinations of higher-order curvature invariants, which successfully produced non-singular homogeneous and isotropic cosmological models. Nevertheless, the method for extending these results to anisotropic singularities or those located within black hole interiors remains an open question.\\

\noindent Since the inclusion of a potential term doesn't help in avoiding the singularity in the context of a Kasner universe or a black hole interior, we shall proceed under the assumption that there is no potential term. This is done by imposing invariance under the shift symmetry. Instead, the authors in \cite{cosmosingu} added only functions of $\Box \phi$ to the Einstein action, 
\begin{equation*}
    \Box \phi \equiv \frac{1}{\sqrt{-g}} \partial_{\mu} \left( \sqrt{-g} g^{\mu \nu} \partial_{\nu} \phi\right) = g^{\mu \nu} \nabla_{\mu} \nabla_{\nu} \phi
\end{equation*}
which ensures invariance under $\phi \rightarrow \phi + C$. Hence, consider the action
\begin{align}
    S = \int \sqrt{-g} d^4x \left( -\frac{1}{2} R + \lambda \left(g^{\mu \nu} \partial_{\mu}\phi \partial_{\nu}\phi - 1 \right) + f (\chi) + L_m \right),
    \label{actionboxphi}
\end{align}
where $8 \pi G$ is set equal to one, $\chi = \Box \phi$, and $L_m$ is the usual matter Lagrangian. The $f(\Box \phi)$ term won't lead to higher derivatives and ghost degrees of freedom due to the mimetic scalar field $\phi$ satisfying the constraint given by equation (\ref{constraint}) (repeated here)
\begin{align*}
    g^{\mu \nu} \partial_{\mu} \phi \partial_{\nu} \phi = 1.
\end{align*}
Upon varying the action with respect to the metric, we get the following equations of motion,
\begin{align}
    {}& G_{\mu \nu} = R_{\mu \nu} - \frac{1}{2} g_{\mu \nu} R =  \tilde{T}_{\mu \nu} + T_{\mu \nu},
    \label{eqmotion}
\end{align}
where $T_{\mu \nu}$ is the usual matter energy-momentum tensor and $\tilde{T}_{\mu \nu}$ is given by
\begin{align}
    {}& \tilde{T}_{\mu \nu} = 2 \lambda \partial_{\mu} \phi  \partial_{\nu} \phi + g_{\mu\nu} \left( \Box \phi f' - f + g^{\rho \sigma} \partial_{\rho} f' \partial_{\sigma} \phi \right) - \left( \partial_{\mu} f' \partial_{\nu} \phi + \partial_{\nu} f' \partial_{\mu} \phi \right),
    \label{tmunutilde}
\end{align}
and $f'$ denotes the derivative of f with respect to $\chi = \Box\phi$. The $\tilde{T}_{\mu \nu}$ term describes the extra contribution to the Einstein equations. In the synchronous coordinate system, the metric is given by \cite{landau}
\begin{equation*}
    g_{00} = 1, g_{0i} = 0, g_{ij} = - \gamma_{ij} (x,t).
\end{equation*}

\subsection{Resolving the Singularity in Friedmann and Kasner Universes}

Considering the flat Friedmann and Kasner universes, where the metric components are functions of time $(\gamma_{ij} = \gamma_{ij}(t))$, the components of the curvature are given by 
\begin {align}
{}& R^0_0 = - \frac{1}{2}\dot{\kappa} - \frac{1}{4}\kappa_i^j \kappa_j^i \label{R00}, \\
& R^i_j = - \frac{1}{2 \sqrt{\gamma}} \frac{d \left( \sqrt{\gamma} \ \kappa^i_j\right)}{dt}, \label{Rij}
\end{align}
where $\gamma = det \ \gamma_{ij}$ and the extrinsic curvature is 
\begin{equation*}
    \kappa^{i}_{j} = \gamma^{im} \dot{\gamma}_{mk},
\end{equation*}
and 
\begin{equation*}
    \kappa = \frac{\dot{\gamma}}{ \gamma}.
\end{equation*}
The $\tilde{T}_{\mu \nu}$ equation given by \eqref{tmunutilde} reduces in the synchronous gauge to
\begin{align*}
    {}& \tilde{T}^0_0 = 2 \lambda + \Box \phi \ f' - f - \dot{\Box}\phi \ f'', \\
    & \tilde{T}^i_j = \left( \Box \phi \ f' - f + \dot{\Box} \phi \ f'' \right) \delta^i_j,
\end{align*}
where the coordinate-independent invariant $\Box \phi$ becomes
\begin{align}
    \chi = \Box \phi = \frac{1}{\sqrt{-g}} \frac{\partial}{\partial x^{\mu}} \left( \sqrt{-g} g^{\mu \nu} \frac{\partial \phi}{\partial x^{\nu}}\right) = \frac{\dot{\gamma}}{2 \gamma} = \frac{\kappa}{2},
    \label{boxphieq}
\end{align}
upon considering the most general solution of the constraint equation \eqref{constraint}, if no coordinate singularities arise,
\begin{equation*}
    \phi = \pm t + A
\end{equation*}
where $A$ is a constant of integration. We see that the function $f$ contributes non-trivially to both $\tilde{T}^0_0$ and $\tilde{T}^i_j$. $f(\chi)$ enables the introduction of the metric and its first derivative in a fully covariant manner when attempting to find a simple modification of General Relativity that avoids singularities.\\

\noindent The $0-0$ components of the equations of motion \eqref{eqmotion}
\begin{equation*}
    R^0_0 - \frac{1}{2} R = \tilde{T}^0_0 + T^0_0,
\end{equation*}
will then take the form
\begin{equation}
    \frac{1}{8} \left( \kappa^2 - \kappa_k^i \kappa_i^k \right) = 2\lambda + \Box \phi f' - f - \dot{\Box \phi} f'' + T^0_0.
    \label{00equation}
\end{equation}
The space-space equation
\begin{equation*}
    R^i_k = \tilde{T}^i_k - \frac{1}{2} \tilde{T} \delta^i_k + T^i_k - \frac{1}{2} T \delta^i_k,
\end{equation*}
gives
\begin{equation}
    \frac{1}{2 \sqrt{\gamma}} \frac{\partial (\sqrt{\gamma} \kappa^i_k)}{\partial t} = (\lambda + \Box \phi f' - f) \delta^i_k - T^i_k + \frac{1}{2} T \delta^i_k.
    \label{spaceeq}
\end{equation}
where $\tilde{T} = \tilde{T}^\alpha_\alpha$ and $T = T^\alpha_\alpha$. These equations of motion are given in terms of the Lagrange multiplier $\lambda$. To determine $\lambda$, we vary the action (eq. \ref{actionboxphi}) with respect to $\phi$. This will give
\begin{equation}
    \frac{1}{\sqrt{\gamma}} \partial_0 \left( 2 \sqrt{\gamma} \lambda \right) = \Box f' = \frac{1}{\sqrt{\gamma}} \partial_0 \left( \sqrt{\gamma} f'' \ \dot{\Box} \phi \right).
    \label{phiequation}
\end{equation}
Upon integrating the above equation, we obtain $\lambda$
\begin{equation}
    \lambda = \frac{C}{2 \sqrt{\gamma}} + \frac{1}{2} f'' \ \dot{\Box} \phi,
    \label{lambdaeq}
\end{equation}
where the integration constant $C$ represents the density of the mimetic fluid.\\

\noindent Assume the energy-momentum tensor for ordinary matter is given by
\begin{equation}
T^{k}{}{i} = -p,\delta^{k}{}{i}.
\end{equation}
This choice is appropriate for the models analyzed in this section and provides sufficient generality to investigate how the presence of matter influences singularities in general spacetimes. In this case, by subtracting one-third of the trace from equation \eqref{spaceeq}, it follows that
\begin{equation}
    \frac{\partial}{\partial t} \left( \sqrt{\gamma} \left( \kappa^i_k - \frac{1}{3} \kappa \delta^i_k \right) \right) = 0,
    \label{subtracting}
\end{equation}
and therefore
\begin{equation}
    \kappa^i_k = \frac{1}{3} \kappa \delta^i_k + \frac{\lambda^i_k}{\sqrt{\gamma}},
    \label{kappaeq}
\end{equation}
where $\lambda^i_k$ are constants of integration which satisfy $\lambda^i_i = 0$. Using (\ref{boxphieq}), (\ref{lambdaeq}), and (\ref{kappaeq}), the $0-0$ equation of motion (\ref{00equation}) will then take the form
\begin{equation}
    \frac{1}{3} (\Box \phi)^2 + f - \Box \phi \ f' = \frac{\lambda^i_k \lambda^k_i}{8\gamma} + \frac{C}{\sqrt{\gamma}} + T^0_0.
    \label{eqfinal}
\end{equation}
In \cite{cosmosingu}, it was argued that to ensure the contributions of $f(\chi)$ do not interfere with General Relativity at low curvatures, the function $f$ can be chosen to be of the Born-Infeld type \cite{BornInfeld}
\begin{equation}
    f(\chi) = 1 + \frac{1}{2} \chi^2 - \chi \arcsin \chi - \sqrt{1 - \chi^2},
\end{equation}
where
\begin{align*}
    f(\chi) \Big|_{\chi=0} = 0, \ f'(\chi) \Big|_{\chi=0} = 0, \  f''(\chi) \Big|_{\chi=0} = 0,
\end{align*}
and $\chi^2 \leq 1$. The term $\frac{1}{2} \chi^2 - \chi \ \arcsin \chi$ was chosen in a way to remove the $\chi^2$ terms in the Taylor expansion of $f(\chi)$. Plus, it removes the singularity in $\frac{df}{d \chi}$ at $\chi = 1$. We introduce a limiting curvature $\chi_m$ so that $\chi \leq \chi_m$. After rescaling $\chi \rightarrow \sqrt{\frac{2}{3}} \frac{\chi}{\chi_m}$ and $f \rightarrow \chi_m^2 f$, $f$ takes the form
\begin{align}
    f(\chi) =  \chi_m^2 \left(1 + \frac{1}{3} \frac{\chi^2}{\chi_m^2} - \sqrt{\frac{2}{3}} \frac{\chi}{\chi_m} \  \arcsin \left(\sqrt{\frac{2}{3}} \frac{\chi}{\chi_m} \right) - \sqrt{1 - \frac{2}{3} \frac{\chi^2}{\chi_m^2}} \right).
\label{funcf}
\end{align}
Using $\Box \phi = \frac{\dot{\gamma}}{2 \gamma}$, equation (\ref{eqfinal}) can be rewritten as 
\begin{equation}
    \frac{1}{12} \left( \frac{\dot{\gamma}}{\gamma}\right)^2 = \epsilon \left( 1 - \frac{\epsilon}{\epsilon_m} \right),
\label{eqofmotion}
\end{equation} 
where 
\begin{align}
    \epsilon = \frac{\lambda^i_k \lambda^k_i}{8\gamma} + \frac{C}{\sqrt{\gamma}} + T^0_0
    \label{epsiloneq}
\end{align}
is the effective energy density and $\epsilon_m = 2\chi_m^2$.\\

\noindent For a flat isotropic Friedmann universe with the metric
\begin{align*}
    ds^2 = dt^2 - a^2(t) \ \delta_{ik}  \ dx^i dx^k,
\end{align*}
$\gamma_{ik} = a^2 \delta_{ik}$. This implies that $\gamma = a^6$ and equation (\ref{kappaeq}) will yield $\lambda_k^i = 0$.  Therefore, equation (\ref{epsiloneq}) becomes  
\begin{align*}
    {}& 3 \left( \frac{\dot{a}}{a} \right)^2 = \frac{\epsilon_m}{a^3} \left( 1 - \frac{1}{a^3} \right) \\
    & \Rightarrow a = \left( 1 + \frac{3}{4} \epsilon_m t^2 \right)^{1/3},
\end{align*}
where the scale factor was normalized in such a way to get $\epsilon = \epsilon_m$ at $a = 1$ for mimetic matter, while the contribution from the other matter is omitted ($T^{0}{}_{0} = 0$). The resulting solution for $a$ depicts a contracting universe initially dominated by cold matter for \( t < -\frac{1}{\epsilon_m} \), with the scale factor following the power law \( a \propto t^{\frac{2}{3}} \). The universe then undergoes a smooth bounce in the time interval \( \frac{-1}{\sqrt{\epsilon_m}} < t < \frac{1}{\sqrt{\epsilon_m}} \). In the post-bounce phase, for \( t > \frac{1}{\sqrt{\epsilon_m}} \), it expands according to the standard dust-dominated Friedmann evolution with \( a \propto t^{\frac{2}{3}} \). To analyze the effect of conventional hydrodynamical matter on the bounce, consider a flat universe filled with matter characterized by a constant equation of state \( p = w \epsilon \), while neglecting mimetic matter. The equations of motion will become
\begin{align*}
    {}& 3 \left( \frac{\dot{a}}{a} \right)^2 = \frac{\epsilon_m}{a^3 ( 1+ w)} \left( 1 - \frac{1}{a^3(1 + w)} \right) \\ 
    & \Rightarrow a = \left( 1 + \frac{3}{4} (1 + w)^2 \epsilon_m t^2 \right)^{\frac{1}{3(1 + w)}},
\end{align*}
which is regular at bounce. Hence, this framework resolves the singularity of Freedmann universe at $t=0$. \\

\noindent Now, we focus on the case of a contracting Kasner universe \cite{landau}, which is a homogeneous but anisotropic solution to Einstein's field equations in a vacuum. The line element of the Kasner universe is given by
\begin{align*}
  ds^2 = dt^2 - t^{2p_1} dx^2 - t^{2p_1} dy^2 - t^{2p_3} dz^2  
\end{align*}
where the constants $p_i$ satisfy
\begin{align*}
    p_1 + p_2 + p_3 = 1,
\end{align*} 
and
\begin{align*}
   p_1^2 + p_2^2 + p_3^2 = 1.
\end{align*} 
For $t < 0$, the metric corresponds to a contracting universe, whereas for $t > 0$, it describes an expanding, homogeneous, and anisotropic universe. For the solution at hand, the scalar curvature and the square of the Ricci tensor both vanish, such that
\[
R = 0
\quad \text{and} \quad
R_{\alpha\beta} R^{\alpha\beta} = 0.
\]
Despite this, the spacetime remains curved because of the non-zero Riemann curvature squared invariant, which is given by
\begin{align*}
     R_{\alpha \beta \gamma \delta} R^{\alpha \beta \gamma \delta} = -\frac{16}{t^4} p_1 p_2 p_3.
\end{align*}
This becomes infinite at t = 0, indicating the presence of a final singularity in a contracting universe and a singularity at the beginning of expansion. We will now explain how, in the proposed theory, this singularity is avoided by limiting the curvature, leading to a bouncing Kasner solution. To find this solution, we consider the metric
\begin{align*}
   \gamma_{ik} = \gamma_{(i)}(t) \delta_{ik},
\end{align*}
that has a determinant equals to $\gamma = \gamma_{(1)} \gamma_{(2)} \gamma_{(3)}$. In an empty universe, equation (\ref{eqofmotion}) becomes
\begin{equation}
    \left( \frac{\dot{\gamma}}{\gamma}\right)^2 = \frac{3 \lambda^i_k \lambda^k_i}{2 \gamma}  \left( 1 - \frac{\lambda^i_k \lambda^k_i}{8\epsilon_m \gamma} \right),
\label{eqofmotionkasner}
\end{equation}
which, upon using equation (\ref{epsiloneq}), becomes \begin{equation*}
    \epsilon = \frac{\lambda^i_k \lambda^k_i}{8\gamma}.
\end{equation*}
Integrating equation (\ref{eqofmotionkasner}) will give us $\gamma$ as a function of time:
\begin{align*}
    \gamma = \frac{\lambda^i_k \lambda^k_i}{8 \epsilon_m} \left( 1 + 3 \epsilon_m t^2\right).
\end{align*}
Therefore, the metric determinant remains finite and possesses a lower bound. The specific components of the metric are determined as follows
\begin{equation}
    \gamma_{(i)} = \left( \frac{\bar{\lambda}^2}{8 \epsilon_m} 
    \left(1 + 3 \epsilon_m t^2 \right) \right)^{1/3} 
    \exp \left( 2 \sqrt{\tfrac{2}{3}} \, \frac{\lambda_{(i)}}{\bar{\lambda}} 
    \sinh^{-1} \!\left(\sqrt{3 \epsilon_m} \, t \right) \right).
\label{kasnermetriccoor}
\end{equation}
This solution is non-singular and has bounded and regular curvature invariants. For $\epsilon_m t^2 \gg 1$, the metric components become 
\begin{equation*}
    \gamma_{(i)} \simeq \left( \frac{\bar{\lambda}^2}{32 \varepsilon_m} \right)^{1/3} \left( 12 \varepsilon_m t^2 \right)^{p_i^{\pm}},
\end{equation*}
where 
\begin{equation*}
    p_i^{\pm} = \frac{1}{3} \pm \sqrt{\frac{2}{3} \frac{\lambda_{(i)}}{\bar{\lambda}}},
\end{equation*}
and $\lambda_{(i)} = (\lambda_1, \lambda_2, \lambda_3)$ are the eigenvalues of $\lambda_k^i$. This solution closely resembles the Kasner solution, with minor higher-order adjustments. For example, the scalar curvature isn't exactly zero at large $t$; instead, it is of the order
\begin{equation*}
    R \sim \frac{1}{\epsilon_m} R_{\alpha\beta\gamma\delta}R^{\alpha\beta\gamma\delta},
\end{equation*}
which is negligible at low curvatures, unless $\epsilon_m$ is comparable to the Planckian value. The Kasner contraction behavior changes only when the curvature approaches the order of $\epsilon_m$. When $\epsilon_m t^2 \ll 1$, the metric (eq. \ref{kasnermetriccoor}) becomes
\begin{equation*}
    \gamma_{(i)} \approx \left(\frac{\bar{\lambda}^2}{8\epsilon_m}\right)^{1/3} \left(1 + \frac{\lambda_{(i)}}{\bar{\lambda}}\sqrt{8\epsilon_m t}\right)
\end{equation*}
describes a smooth bounce, leading to an expanding Kasner universe.\\

\noindent In this framework, we saw that the singularities in cosmological solutions can be resolved and that the contracting universes can undergo a bounce at the limiting curvature. All curvature invariants remain regular and bounded by values determined by $\epsilon_m$. The bounce occurs within a short time interval \( t \sim \epsilon_m^{-1/2} \), with general relativity recovering outside this window. The introduction of a mimetic field that modifies gravity with Born-Infeld-type corrections naturally provides a viable dark matter candidate while significantly altering Einstein’s equations only at very high curvatures. Although a specific function $f(\chi)$ was chosen for simplicity, other functions that contain a Born-Infeld term and have a finite derivative $df/d\chi$ as $\chi$ approaches its limit will work.  \\

\noindent The above considerations are limited to highly symmetric spacetimes. One might wonder whether the curvature remains generically bounded in arbitrary inhomogeneous spaces. In \cite{de2020singularity}, it was investigated whether introducing a mimetic field into Kantowski-Sachs cosmology can resolve the initial singularity, while in \cite{gorji2018higher}, higher derivative mimetic gravity was explored. It was demonstrated how the addition of higher-order derivatives to the mimetic gravity framework can lead to new cosmological solutions and to mimetic nonsingular bouncing scenarios. Below, we will demonstrate how the singularity can also be avoided in the case of a black hole.

\subsection{Non-singular Black Holes} 

The issue of singularities within black holes has, for a long time, constituted one of the most profound and persistent challenges in theoretical physics. In this subsection, we will see how this singularity is dealt with in the context of mimetic gravity with limiting curvature \cite{nonsingularblack}. In the previous subsection, it was shown that the introduction of limiting curvature prevents the curvature from becoming infinite. Consequently, we will see how the central singularity of the Schwarzschild black hole can be removed. \\

\noindent We start from the action introduced above (equation \ref{actionboxphi}) without the usual matter
\begin{align*}
    S = \int \sqrt{-g} \ d^4x \left( -\frac{1}{2} R + \lambda \left(g^{\mu \nu} \partial_{\mu}\phi \partial_{\nu}\phi - 1 \right) + f (\chi)\right),
\end{align*}
and we are working in a synchronous coordinate system with the metric given by 
\begin{align*}
    ds^2 = dt^2 - \gamma_{ik} \left(t,x^l\right) dx^i dx^k.
\end{align*}
Upon varying the action with respect to the metric $g^{\mu \nu}$, we get equation (\ref{eqmotion}) with the $T_{\mu \nu}$ term equals to zero and the $\tilde{T}_{\mu \nu}$ given by equation (\ref{tmunutilde}). The time-time component of the curvature is still given by equation (\ref{R00}), while the space-space components are given by \cite{landau}, \cite{nonsingularblack}
\begin{align*}
& R^i_j = - \frac{1}{2 \sqrt{\gamma}} \frac{d \left( \sqrt{\gamma} \kappa^i_j\right)}{dt} - P^i_j,
\end{align*}
where $P^i_j$ is the three-dimensional Ricci tensor for the metric $\gamma_{ij}$. The $0-0$ and the space-space equations of motion now take the forms (similar to equations (\ref{00equation}) and (\ref{spaceeq}))
\begin{equation}
    \frac{1}{8} \left( \kappa^2 - \kappa_k^i \kappa_i^k + 4P \right) = 2\lambda + \chi f' - f - \dot{\chi} f'',
\label{00eq2}
\end{equation} 
and
\begin{equation}
    \frac{1}{2 \sqrt{\gamma}} \frac{\partial (\sqrt{\gamma} \kappa^i_k)}{\partial t} + P^i_k = (\lambda + \chi f' - f) \delta^i_k.
\label{spaceeq2}
\end{equation}
Variation of the action with respect to $\phi$ gives (compare to equation (\ref{phiequation}))
\begin{equation*}
    \frac{1}{\sqrt{\gamma}} \partial_0 \left( 2 \sqrt{\gamma} \lambda \right) = \Box f' = \frac{1}{\sqrt{\gamma}} \partial_0 \left( \sqrt{\gamma} f'' \ \dot{\Box} \phi \right) - \triangle f',
\end{equation*}
where $\triangle f'$ is the covariant Laplacian of $f'$ for the metric $\gamma_{ik}$. In a very similar manner to what was done previously in this section, the above equation can be used to determine the Lagrange multiplier $\lambda$. Assuming that the determinant of the metric is factorizable, $\gamma(t, x^i) = \gamma_1(t) \gamma_2 (x^i)$, then both $\chi$ and $\kappa$ depend only on time and $\triangle f'$ does not contribute as it vanishes. Therefore, the $\phi$-equation given above reduces to (\ref{phiequation}), and the Lagrange multiplier is still given by equation (\ref{lambdaeq}). We will set the constant of integration $C$, which represents mimetic cold matter, equals to zero $(C = 0)$ because the latter behaves exactly like dust, and it was shown above that the ordinary matter does not have a substantial impact on resolving anisotropic singularities. The analogous of equation (\ref{subtracting}), which is obtained by subtracting from equation (\ref{spaceeq2}) one third of its trace, is given by
\begin{equation*}
    \frac{\partial}{\partial t} \left( \sqrt{\gamma} \left( \kappa^i_k - \frac{1}{3} \kappa \delta^i_k \right) \right) = -2 \left( P^i_k - \frac{1}{3} P \delta^i_k \right) \sqrt{\gamma},
\end{equation*}
from which we can get $\kappa^i_k$
\begin{align*}
    \kappa^i_k = \frac{1}{3} \kappa \delta^i_k + \frac{\lambda^i_k}{\sqrt{\gamma}},
\end{align*}
where $\lambda^i_k$ is traceless $(\lambda^i_i = 0)$ and given by 
\begin{equation*}
\lambda^i_k = -2 \int \left( P^i_k - \frac{1}{3} P \delta^i_k \right) \sqrt{\gamma} \, dt.
\end{equation*}
Using the above, we can rewrite the $0-0$ equation of motion (\ref{00eq2}) as
\begin{equation*}
\frac{1}{12} \chi^2 + f - \chi f' = \frac{\lambda^i_k \lambda^k_i}{8\gamma} - \frac{1}{2} P,
\end{equation*}
which will be given in a particularly simple equation, after substituting the function $f$ from equation (\ref{funcf}),
\begin{equation}
\chi_m^2 \left( 1 - \sqrt{1 - \frac{2}{3} \frac{\chi^2}{\chi_m^2}} \right) = \varepsilon,
\label{varepsilon}
\end{equation}
where
\begin{equation*}
\varepsilon = \frac{\lambda^i_k \lambda^k_i}{8\gamma} - \frac{1}{2} P
\end{equation*}
is independent of the time derivative of the metric. Using $\chi = \frac{\dot{\gamma}}{2 \gamma}$, we can rewrite equation (\ref{varepsilon}) as
\begin{equation}
\frac{1}{12} \left( \frac{\dot{\gamma}}{\gamma} \right)^2 = \varepsilon \left( 1 - \frac{\varepsilon}{\varepsilon_m} \right).
\label{mastereq}
\end{equation}
The equation above will be used to analyze the black hole solution. \\

\noindent The Schwarzschild metric is given by
\begin{equation*}
    ds^2 = \left(1 - \frac{r}{r_g} \right) dt_S^2 - \frac{dr^2}{\left(1 - \frac{r}{r_g}\right)} - r^2  \left(d \theta^2 + sin^2 \theta d \phi^2 \right),
\end{equation*}
where $r_g$ is the gravitational radius. As shown in \cite{nonsingularblack}, inside the Schwarzschild black hole, the metric can rewritten as
\begin{equation}
    ds^2 = dt^2 - a^2(t) dR^2 - b^2(t) \left(d \theta^2 + sin^2 \theta d \phi^2 \right),
\label{synch}
\end{equation}
where 
\begin{equation}
    a^2(t) = \frac{1 - \tau^2(t)}{\tau^2(t)}, \qquad b^2(t) = \tau^4(t) r_g^2,
\label{a&b}
\end{equation}
and 
\begin{equation*}
    t = r_g \left(arcsin \ \tau - \tau \sqrt{1 - \tau^2} \right).
\end{equation*}
The synchronous coordinate system (\ref{synch}) is initially chosen due to its convenience in finding a nonsingular generalization of the Schwarzschild solution in the theory with limiting curvature. However, it is not free of coordinate singularities because $\gamma = \left( 1 - \tau^2\right) \tau^6 r_g^4$ vanishes as $\tau^2 \rightarrow 1$, and $\phi = t + A$ should be modified. To overcome these issues and find a synchronous coordinate system free of fictitious singularities, a coordinate transformation from $(t, R)$ to $(T, \bar{R})$ is introduced, defined by 
\begin{align}
T &= R + \int \frac{\sqrt{1+ {a}^2}}{a} dt, \\ \nonumber
\bar{R} &= R + \int \frac{dt}{a\sqrt{1+ {a}^2}}.
\label{coord_transform} 
\end{align}
This new coordinate system aims to provide a more robust framework for analyzing the near-horizon geometry. The metric (\ref{synch}) can be rewritten in new synchronous coordinates $(T, \bar{R})$ for the Schwarzschild solution (\ref{a&b})
\begin{equation}
    ds^2 = dT^2 - \tau^{-2} d \bar{R}^2 - \tau^4 r_g^2 d \Omega^2,
\end{equation}
where 
\begin{equation*}
    T - \bar{R} = \frac{2}{3} r_g \tau^3.
\end{equation*}
In the case of the Schwarzschild solution, this leads to the Lemaître coordinate representation, which remains regular at the horizon and accounts for both the interior and exterior regions of the black hole. The corrected solution for $\phi$ is then given by
\begin{equation}
    \phi = T = R + \int \frac{\sqrt{1+{a}^2}}{a} dt.
\label{phi_corrected}
\end{equation}
Despite the regularity of Lemaître coordinates, the non-separability of space and time coordinates in the metric components makes them inconvenient for probing the internal structure of nonsingular black holes. The analysis returns to the original system (equation \ref{synch}) but using the correct solution for $\phi$ (equation \ref{phi_corrected}). The latter satisfies the constraint equation (\ref{constraint}). Calculating $\chi = \Box\phi$ yields
\begin{equation}
    \chi = \Box\phi = \frac{\dot{\gamma}}{2\gamma} \sqrt{1+\frac{1}{{a}^2}} + \frac{d}{dt} \sqrt{1+\frac{1}{{a}^2}}.
\label{chi_general}
\end{equation}
For a Schwarzschild black hole, this simplifies to
\begin{equation}
    \chi = \frac{3}{2r_g \tau^3},
\label{chi_schwarzschild}
\end{equation}
and on the horizon we have $\chi^2 \ll \epsilon_m$ for $r_g \gg \epsilon_m^{-1/2}$. In \cite{nonsingularblack}, it was argued that $\chi$ can be set as
\begin{equation*}
    \chi = \frac{\dot{\gamma}}{2 \gamma},
\end{equation*}
with good accuracy, and then equation (\ref{mastereq}) can be used to investigate how and what happens when we approach the limiting curvature and beyond. It was shown that applying the principle of limiting curvature to the Schwarzschild black hole results in the removal of its central singularity and yields a geodesically complete spacetime where infalling trajectories do not terminate abruptly. This means that the paths (geodesics) of objects falling into the black hole do not abruptly end at a singularity but can be continued. The resulting internal structure is intriguingly described using a ``Russian nesting dolls" analogy: an observer falling past the event horizon (radius $r_g$) does not encounter a singularity; they enter a region where the curvature reaches its maximum allowed value (the limiting curvature). After briefly traversing this high-curvature zone, the observer re-emerges into a spacetime locally resembling the near-horizon region of a smaller effective Schwarzschild black hole, with a gravitational radius scaling proportionally to ($r_g^{(1/3)}$). This process repeats iteratively, with each passage through the limiting curvature region leading to emergence into an even smaller effective Schwarzschild interior (radii scaling as ($r_g^{(1/9)}$, then $r_g^{(1/27)}$, and so on). Ultimately, after a finite number of cycles, the observer reaches a final state, remaining indefinitely within the region characterized by the limiting curvature. They do not re-emerge into a smaller Schwarzschild region after this point. In essence, this theoretical framework replaces the point-like singularity with a ``bounce" mechanism mediated by limiting curvature, creating a nested, self-similar internal structure where observers transition through progressively smaller effective Schwarzschild geometries before reaching a final, non-singular state.\\

\noindent In this section, we saw that the modified mimetic gravity, with a $f(\Box \phi)$ term, offers a potential resolution of both cosmological and black hole singularities \cite{cosmosingu}, \cite{nonsingularblack}. The core concept of this approach is quite straightforward. In a synchronous coordinate system with the metric $ds^2 = -dt^2 + \gamma_{ij} dx^i dx^j$ and a scalar field $\phi = \pm t + const$, the d'Alembertian of $\phi$ becomes $\chi \equiv \Box \phi = \frac{\dot{\gamma}}{2 \gamma}$. As one approaches the singularity, the argument of the function f grows without bound. By selecting a limiting value $\chi_m$, ideally much lower than the Planck scale to avoid quantum gravitational effects, one can tune the function $f$ by adopting a Born-Infeld-type form, such that $f$ stabilizes as $\chi \rightarrow \chi_m$. This adjustment has been shown to yield a bounce-like resolution of singularities.\\

Exploring the non-singular black holes within the framework of the Limiting Curvature Mechanism, from a Hamiltonian perspective, was done in \cite{achour2018non}. By solving the equations of motion in the regime deep inside the black hole, a black hole with no singularity was recovered, due to the limiting curvature mechanics. However, further analysis of the above LCM model was explored in \cite{golovnev2}. It was revealed that the non-singular bounce depends critically on a subtle branch-changing mechanism for the multi-valued function $f$, which has been specifically tuned to vanish rapidly for small arguments. This mechanism not only enables the bounce but has also attracted considerable attention due to its similarities with loop quantum cosmology \cite{LQC1, LQC2}. However, it was shown that a bounce is impossible within the model's `trivial branch'. This impossibility arises from the fact that the conditions required for $f$ at the bounce point $(H=0)$, namely $f(0) = f'(0) = f''(0) = 0$, inevitably lead to $\dot{H}=0$, thus excluding the necessary cosmic turnaround \cite{Firouzjahi}.\\

Other work has also investigated black hole solutions in various formulations of mimetic gravity. For instance, the authors of \cite{chen2018black} explored black hole solutions within mimetic Born-Infeld gravity, aiming to find regular black holes where the central singularity is resolved due to the Born-Infeld terms, while the mimetic scalar field influences the spacetime geometry. Reference \cite{nashed2019charged} investigates charged, rotating black hole solutions within mimetic gravity, where the electromagnetic field is described by nonlinear electrodynamics. 

\section{Instabilities in Mimetic Matter Perturbations}
\label{section7}

In \cite{Firouzjahi}, the authors investigate the perturbative stability of generalized mimetic gravity models where the action is supplemented by a general higher-derivative function $f(\chi)$, with $\chi \equiv \Box\phi$. This modification is primarily motivated by the need to generate a non-zero sound speed for scalar perturbations, which is required to prevent the formation of caustic singularities and to allow for a well-defined quantum treatment of fluctuations. However, the central finding of the paper is that these models are generically plagued by gradient instabilities, characterized by an imaginary sound speed $(c_s^2 < 0)$ in the scalar sector at the linear perturbative level.\\

\noindent The theory studied is defined by the action
\begin{equation}
S = \int d^4x \sqrt{-g} \left[ \frac{M_P^2}{2} R + \lambda \left( g^{\mu\nu}\partial_\mu\phi\partial_\nu\phi + 1 \right) + f(\chi) - V(\phi) \right].
\end{equation}
On a spatially flat Friedmann--Robertson--Walker (FRW) background, the mimetic constraint fixes the background field to $\phi = t$. The background Einstein equations determine the value of the Lagrange multiplier as
\begin{equation}
\lambda = (M_P^2 - 3f_{\chi\chi})\dot{H} = \epsilon (3f_{\chi\chi} - M_P^2)H^2 ,
\end{equation}
where $f_{\chi\chi} \equiv d^2f/d\chi^2$ and $\epsilon \equiv -\dot{H}/H^2$.\\

\noindent The stability analysis is first performed in the comoving gauge ($\delta\phi=0$). By expanding the action to quadratic order (setting $M_P=1$), the authors derive the quadratic action for the curvature perturbation $\mathcal{R}$
\begin{equation}
S^{(2)}_{\text{com}} = \int dt\, d^3x\, a^3 \left[ \left(3 - \frac{2}{f_{\chi\chi}}\right)\dot{\mathcal{R}}^2 + \frac{(\partial \mathcal{R})^2}{a^2} \right].
\end{equation}
This yields a sound speed for scalar perturbations given by
\begin{equation}
c_s^2 = \frac{f_{\chi\chi}}{2 - 3 f_{\chi\chi}} .
\end{equation}
While $c_s^2$ can be rendered non-zero and positive by choosing an appropriate $f(\chi)$, the corresponding Hamiltonian density in Fourier space reveals a fundamental pathology
\begin{equation}
\mathcal{H}_{\text{com}} = - \frac{c_s^2}{4a^2}\Pi_{\mathcal{R}}^2 - a^2 k^2 \mathcal{R}^2 .
\end{equation}
The authors show that for $c_s^2 > 0$, the kinetic term has the wrong sign (a ghost), whereas for $c_s^2 < 0$, the theory suffers from gradient instabilities. Crucially, they conclude that the Hamiltonian is never bounded from below regardless of the choice of $f(\chi)$, as both the kinetic and gradient terms carry signs that lead to instability.\\

\noindent The analysis is repeated in the Newtonian gauge to ensure gauge invariance of the physical results. In this gauge, the metric perturbations are related to the mimetic field via the constraint $\Phi = \delta\dot{\phi}$. The resulting equation of motion for $\delta\phi$ is
\begin{equation}
\ddot{\delta\phi} + H\dot{\delta\phi} + \dot{H}\delta\phi - c_s^2\frac{\partial^2\delta\phi}{a^2} = 0 ,
\end{equation}
consistent with the comoving gauge result. A detailed Hamiltonian treatment in this gauge confirms that the mimetic constraint protects the theory from Ostrogradsky ghosts (extra degrees of freedom); however, the single existing degree of freedom remains dynamically unstable. The Hamiltonians in both gauges are shown to be related by a canonical transformation, confirming the physical consistency of the instability.\\

\noindent The authors also conduct a full Dirac constraint analysis without fixing the gauge or imposing the mimetic constraint by hand at the perturbative level. This rigorous approach demonstrates that the mimetic condition $\dot{\phi}=1$ is recovered as a consistency relation (a secondary constraint). The full analysis confirms that the theory propagates exactly one scalar degree of freedom, which possesses a Hamiltonian that is not bounded from below. Thus, while higher-derivative terms of the form $f(\Box\phi)$ successfully generate a non-zero sound speed, they inevitably render the mimetic matter scenario perturbatively unstable.

\section{Ghost Free Mimetic Massive Gravity}
\label{section8}

The simplest formulation of massive gravity within General Relativity, aiming to preserve diffeomorphism invariance without explicit breaking, often employs a Brout-Englert-Higgs (BEH) mechanism. This typically involves introducing four scalar fields, $\phi^A$ ($A = 0, 1, 2, 3$) \cite{BEH1, BEH2, BEH3}. These scalar fields are assumed to acquire vacuum expectation values in spacetime, given by
\begin{equation}
    \langle\phi^A\rangle = \delta^A_{\mu} x^{\mu}.
\label{eq:vev_phi}
\end{equation}
With a Higgs-type potential, expanding the fields as $\phi^A = x^A + \chi^A$ leads to three of the four scalars being absorbed by the graviton, thereby giving it a mass. The remaining scalar, $\chi^0$, typically introduces a ghost instability, unless the mass term in the Lagrangian is chosen to be of the Fierz–Pauli form \cite{fierzpauli}, in which case the ghost mode is absent at the linearized level. Nevertheless, nonlinear interactions or nontrivial backgrounds reintroduce the Boulware–Deser ghost \cite{BDg}. An enormous amount of work was done to find a form of the action where the ghost state is not excited. A proposal by Gabadadze and collaborators \cite{gabadadze}, known as dRGT, showed that the ghost state decouples in a time-independent background, resulting in a theory whose action involves an infinite expansion of a square root function. The action is given by 
\begin{equation}
    S = \int d^4x \sqrt{-g} \left[ -\frac{1}{2}R + \frac{m^2}{2} \left( S^2 - S^{AB}S_{AB} \right) \right], 
\label{massive_gravity_action}
\end{equation}
where $S_{AB}$ is defined in terms of the metric perturbation $\bar{h}_{AB}$:
\begin{equation}
    S_{AB} = \sqrt{\eta_{AB} + \bar{h}_{AB}} - \eta_{AB}.
\label{S_AB_definition}
\end{equation}
Here, the square root denotes a matrix square root. At second order in the metric perturbation $\bar{h}$, this theory \eqref{massive_gravity_action} reduces to the standard Fierz-Pauli theory. However, at higher orders, the theory is represented by an infinite series in powers of $\bar{h}$. It is through the introduction of auxiliary fields, often of the vierbein type, that the square root structure appearing in the definition of $S_{AB}$ can be expressed in a finite, closed form, leading to a theory that becomes quadratic in the auxiliary field $S$ \cite{quadraticaction}. Despite these efforts, concerns about the theory's consistency persist. While the action presented in Eq. \eqref{massive_gravity_action} appears promising from the perspective of keeping the scalar field perturbation $\chi^0$ non-dynamical to all orders, subsequent investigations, such as those in references \cite{exorcisingtheghost, hiddenghost}, have shown that certain terms, like $\bar{h}^{0i} \bar{h}_{0i}$, which are already present in the quadratic FP term, can still lead to the excitation of the ghost mode in certain specific background spacetimes. Models avoiding the Boulware-Deser ghost instability were also studied in \cite{creminelli2005ghosts, hassan2012resolving, hassan2012ghost, hassan2012confirmation, hassan2012proof}. \newline

In contrast, making use of the mimetic formulation of gravity, a ghost-free massive gravity can be formulated where the ghost state is completely eliminated. The idea is to use the constrained mimetic field $\phi$ as the $\phi^0$ in $\phi^A$ to give mass to the graviton \cite{ghostfreemassive}. Since this field is always in the broken symmetry phase due to its constraint, the problematic degree of freedom is replaced by dark matter, and the ghost is avoided at all orders in perturbation theory. Then only three scalar fields will be absorbed by the graviton. As we show below, the mass term in this setup must differ from the usual Fierz-Pauli form. The proposed action (not of Fierz-Pauli type) is given by \cite{ghostfreemassive}
\begin{align}
    I = \int d^4 x \sqrt{-g} \left( -\frac{1}{2} R + \frac{m^2}{8} \left( \frac{1}{2} \bar{h}^2 - \bar{h}^{AB} \bar{h}_{AB} \right) + \lambda \left( g^{\mu \nu} \partial_{\mu} \phi^0 \partial_{\nu} \phi^0 - 1\right) \right)
    \label{mimeticmassive}
\end{align}
where $\bar{h}^{AB}$ is a diffeomorphism invariant set of scalars used to give mass to the graviton and is given by 
\begin{equation*}
    \bar{h}^{AB} = g^{\mu \nu} \partial_{\mu} \phi^A \partial_{\nu} \phi^B - \eta^{AB}. 
\end{equation*}
The capital letters will be raised and lowered using the auxiliary Minkowski metric, $\eta^{AB} = (1, -1, -1, -1)$. The last term in the action (\ref{mimeticmassive}) reflects the mimetic origin of $\phi_0$, while the mass term features a relative coefficient of $\frac{1}{2}$ between $\bar{h}^2$ and $\bar{h}_{AB} \bar{h}^{AB}$, differing from Fierz-Pauli term. Varying the action with respect to $\delta \lambda$ clearly gives
\begin{equation*}
\bar{h}^{00} = 0.
\end{equation*}
Variation with respect to $\delta g^{\mu \nu}$ and next with respect to $\delta \phi^A$ will give respectively
\begin{align*}
G_{\mu \nu} = 
& -\frac{m^2}{8} \left( \frac{1}{2} \bar{h}^2 - \bar{h}^{AB} \bar{h}_{AB} \right) g_{\mu \nu} 
+ \lambda \left( 2 \partial_{\mu} \phi^0 \partial_{\nu} \phi^0 \right) \nonumber \\
& + \frac{m^2}{2} \left( \frac{1}{2} \bar{h} \partial_{\mu} \phi^A \partial_{\nu} \phi^A 
- \bar{h}_{AB} \partial_{\mu} \phi^A \partial_{\nu} \phi^B \right),
\end{align*}
and
\begin{equation}
    \nabla^{\mu} \left( 
    m^2 \left( \frac{1}{2} \bar{h} \partial_{\mu} \phi^A - \bar{h}_{AB} \partial_{\mu} \phi^B \right) 
    + 4 \lambda \delta^0_A \partial_{\mu} \phi^0 
    \right) = 0.
\label{delphieq}
\end{equation}
Considering small perturbations around a Minkowski background, the metric $g_{\mu\nu}$ and the set of scalar fields $\phi^A$ are perturbed as
\begin{align*}
    g_{\mu\nu} &= \eta_{\mu\nu} + h_{\mu\nu}, \\
    \phi^A &= x^A + \chi^A.
\end{align*}
The equations of motion are then linearized with respect to $h_{\mu\nu}$ and $\chi^A$, given that $\lambda$ is of first order in perturbations. By first setting $A=0$ in equation (\ref{delphieq}), we get
\begin{equation}
    \partial_0 \lambda - \frac{m^2}{4} \left( \partial^0 \bar{h}_{\rho 0} - \frac{1}{2} \partial_0 \bar{h} \right) = 0,
\label{A=0}
\end{equation}
and subsequently, by setting $A=k$, we obtain
\begin{equation}
    m^2 \left( \partial^0 \bar{h}_{\rho k} - \frac{1}{2} \partial_k \bar{h} \right) = 0.
\label{A=k}
\end{equation}
The linearized Einstein tensor in terms of $h_{\mu\nu}$ is given by
\begin{align}
G_{\mu\nu}(h_{\rho\sigma}) ={}& \frac{1}{2} \left( \partial^2 h_{\mu\nu} - \partial_\mu \partial^\rho h_{\rho\nu} - \partial_\nu \partial^\rho h_{\rho\mu} + \partial_\mu \partial_\nu h \right) \nonumber \\
& + \frac{1}{2} \eta_{\mu\nu} \left( \partial^2 h - \partial^\sigma \partial^\rho h_{\rho\sigma} \right),
\label{Gmunumassive}
\end{align}
where $\partial^2 \equiv \partial^\mu \partial_\mu \equiv \Box$ and $h \equiv \eta^{\mu\nu} h_{\mu\nu}$. To first order in perturbations, the quantity $\bar{h}^{AB}$ is defined as
\begin{equation*}
\bar{h}^{AB} = \delta^A_\mu \delta^B_\nu h^{\mu\nu} + \partial^A \chi^B + \partial^B \chi^A,
\end{equation*}
where $h^{\mu\nu} = g^{\mu\nu} - \eta^{\mu\nu}$, and using the Minkowski metric $\eta_{AB}$ to lower capital indices and then replacing them with Greek indices, we get
\begin{equation*} 
h_{\mu\nu} = -\bar{h}_{\mu\nu} + \partial_\mu \chi_\nu + \partial_\nu \chi_\mu.
\end{equation*}
Substituting the latter into \eqref{Gmunumassive}, the terms involving $\chi$ cancel, yielding $G_{\mu\nu}(h_{\rho\sigma}) = G_{\mu\nu}(-\bar{h}_{\rho\sigma})$. This implies the linearized Einstein tensor can be expressed using the gauge-invariant variables $\bar{h}_{\rho\sigma}$.
With the constraint $\bar{h}_{00}=0$, the linearized Einstein equations are stated as:
\begin{align*}
G_{00}(-\bar{h}_{\rho\sigma}) &= 2\lambda + \frac{m^2}{4}\bar{h},  \\
G_{0i}(-\bar{h}_{\rho\sigma}) &= -\frac{m^2}{2} \bar{h}_{0i}, \\
G_{ij}(-\bar{h}_{\rho\sigma}) &= -\frac{m^2}{2} \left( \bar{h}_{ij} - \frac{1}{2} \eta_{ij} \bar{h} \right). 
\end{align*}
In the standard Fierz-Pauli theories of massive gravity, where we use the mass term, $\bar{h}^2 - \bar{h}^{AB}\bar{h}_{AB}$, Bianchi identities lead to the vanishing of the perturbations of the scalar curvature $\delta R = 0$, and consequently we face the vDVZ discontinuity \cite{vDVZ1}, \cite{vDVZ2}, a problem usually tackled with non-linear corrections \cite{corrections}. However, in this case, the Bianchi identities will impose conditions \eqref{A=0} and \eqref{A=k}. Therefore, the vDVZ discontinuity is already absent at the linear level. \\

\noindent As shown in \cite{ghostfreemassive}, $G_{ij}(-\bar{h}_{\mu\nu})$ can be simplified using equations \eqref{A=0} and \eqref{A=k}, and this will give
\begin{equation*}
\partial^2 \bar{h}_{ij} - \eta_{ij} \left( \frac{1}{2} \partial^2 \bar{h} - \frac{4\dot{\lambda}}{m^2} \right) = -m^2 \left( \bar{h}_{ij} - \frac{1}{2} \eta_{ij} \bar{h} \right),
\end{equation*}
from which immediately follows that the traceless part of spatial metric components,
\begin{equation*} 
\bar{h}_{ij}^T \equiv \bar{h}_{ij} - \frac{1}{3} \eta_{ij} \bar{h},
\end{equation*}
satisfies the wave equation
\begin{equation} \label{massequation}
(\Box + m^2) \bar{h}_{ij}^T = 0,
\end{equation}
which describes the massive graviton with five degrees of freedom. $\bar{h}_{0i}$ and $\bar{h} = \bar{h}^i_i$ are completely determined from the equations of motion in terms of $\bar{h}_{ij}$ and $\lambda$, and $\lambda$ describes mimetic matter satisfying the equation
\begin{align*}
    \ddot{\lambda} + \frac{m^2}{4} \lambda = 0.
\end{align*}
Therefore, we conclude that massive mimetic gravity describes a massive graviton, characterized by the traceless component $\bar{h}^{T}_{ij}$ satisfying equation \eqref{massequation}, along with mimetic matter represented by $\lambda$. The use of this mimetic field, maintained in a broken symmetry phase, mandates a non-Fierz-Pauli mass term characterized by a $-1/2$ relative coefficient between the quadratic invariants—distinct from the standard Fierz-Pauli value of $-1$—to achieve a massive graviton. This specific structure ensures the graviton's decoupling from mimetic matter at the linear level and successfully resolves the van Dam--Veltman--Zakharov (vDVZ) discontinuity. 
As demonstrated in~\cite{beyondlinear}, the $h_{00}$ metric component is constrained and replaced by the mimetic scalar field, which dynamically imitates cold dark matter; this substitution ensures that the theory remains completely ghost-free at all orders, regardless of non-linear extensions.\\

In this section, we have seen that by employing Brout-Englert-Higgs mechanism with four scalar fields, where one is constrained as in mimetic gravity, a consistent massive gravity theory is generated possessing only five degrees of freedom. 
The sixth mode, typically associated with the Boulware-Deser ghost, is constrained and effectively replaced by mimetic matter to all orders in the theory, a feature that also contributes to the resolution of the vDVZ discontinuity. The Hamiltonian formulation of this Ghost-Free Mimetic Massive Gravity theory was rigorously studied in \cite{malaeb2019hamiltonian}, where an analysis of the degrees of freedom proved that only the five physical modes remain. Furthermore, the broader cosmological implications of this mimetic theory of massive gravity were explored in~\cite{solomon2019massive}.

\section{Mimetic Horava Gravity} 
\label{section9}

General relativity is not a renormalizable theory, and it cannot be quantized using standard quantization techniques. An attempt to make it renormalizable is by including higher-order curvature terms that would modify the graviton's propagator at high energies. This was done in 1977, when Stelle \cite{stelle_renormalization} showed that adding curvature-squared terms makes gravity renormalizable, but at the cost of introducing ghosts and violating unitarity due to higher-order time derivatives. One approach involves modifying the propagator by introducing higher-order spatial derivatives while avoiding higher-order time derivatives. This method could potentially improve the theory's ultraviolet (UV) behavior of the graviton propagator. However, this breaks Lorentz invariance, as space and time are treated differently. Although Lorentz invariance is abandoned at very high energies (in the UV), the aim is to recover it at low energies (in the IR), thereby preserving the successes of General Relativity in the classical regime. Ho\v{r}ava constructed such a model of quantum gravity \cite{Horava1, Horava2}, now widely known as Ho\v{r}ava-Lifshitz gravity, in which Lorentz symmetry is explicitly broken, enabling the addition of terms in the action that depend on the spatial Ricci tensor, the curvature scalar, and their spatial derivatives. Furthermore, renormalizability in this model is restricted to its projection onto the $\mathbb{R} \times \Sigma_3$ product space. Restoring covariance by introducing a dynamical scalar or vector field \cite{Barvinsky1, Germani} tends to nullify this attribute. Such extensions involve additional propagating degrees of freedom, which has limited their acceptance as a robust method for achieving renormalizable gravity.\\

\noindent Within the framework of mimetic gravity, the scalar field facilitates the construction of diffeomorphism-invariant models that map to Ho\v{r}ava gravity when using the synchronous gauge \cite{mimetichorava}. This equivalence relies on utilizing $\phi$ as a specific synchronous time coordinate to establish a unique $3+1$ slicing. The gradient of the mimetic field defines a timelike unit vector, $n_{\mu} = \partial_{\mu} \phi$, which aligns with the temporal direction. This vector enables the formulation of a projection operator to map 4D tensors onto 3D spatial slices, thereby dynamically generating the preferred spacetime foliation characteristic of Ho\v{r}ava gravity. This allows one to incorporate the higher-order spatial derivative terms characteristic of Ho\v{r}ava gravity (which ensure power-counting renormalizability and can resolve singularities) into a fully covariant action. Such a projection operator is given by
\begin{equation*}
    P^{\mu}_{\nu} = \delta^{\nu}_{\mu} - \partial_{\mu} \phi \partial_{\kappa}\phi \ g^{\mu \kappa},
\end{equation*}
satisfying 
\begin{align*}
    P_{\mu}^{\rho}P_{\rho}^{\nu} = P_{\mu}^{\nu} \quad P^{\nu}_{\mu} \partial_{\nu} \phi = 0.
\end{align*}
In synchronous gauge, 
\begin{equation*}
    P^0_0 = 0, \ P^i_0 = 0 = P_i^0, \ P^j_i = \delta^j_i,
\end{equation*}
showing that $P_{\mu}^{\nu}$ projects space-time vectors to space vectors. By noting that 
\begin{align*}
    {}& R^{0}_{kij} = D_{i}K_{kj} - D_j K_{ki} \\
    & R^{0}_{k0j} = \dot{K}_{kj} - K_{jn}K^n_{k} \\
    & R^{l}_{kij} = \prescript{(3)}{} R^l_{kij} + K^l_i K_{jk} - K^l_j K_{ik},
\end{align*}
we can construct the following
\begin{equation*}
    \tilde{R}^{\sigma}_{\rho \mu \nu} = P^{\sigma}_{\delta} P^{\gamma}_{\rho} P^{\alpha}_{\mu} P^{\beta}_{\nu} R^{\delta}_{\gamma \alpha \beta} + \nabla_{\mu} \nabla^{\sigma} \phi \nabla_{\rho} \nabla_{\mu} \phi,
\end{equation*}
whose the only non-vanishing components are $\prescript{(3)}{} R^l_{kij}$ in the synchronous gauge. The $\tilde{R}_{\mu \nu}$ tensor can also be defined
\begin{align*}
\tilde{R}_{\mu\nu} := P^\alpha_\mu P^\beta_\nu R_{\alpha\beta} + \Box\phi \nabla_\mu\nabla_\nu\phi - \nabla_\mu\nabla^\rho\phi\nabla_\nu\nabla_\rho\phi - R^\gamma_{\mu\delta\nu} \nabla^\delta\phi\nabla_\gamma\phi.
\end{align*}
It's non-zero components coincide with $\prescript{(3)}{} R_{ij}$, and upon contracting with $g^{\mu \nu}$, we get
\begin{equation*}
\tilde{R} := 2R^{\mu\nu} \partial_\mu\phi\partial_\nu\phi - R - (\Box\phi)^2 + \nabla_\mu\nabla_\nu\phi\nabla^\mu\nabla^\nu\phi.
\end{equation*}
\noindent Consider the action, written in terms of four-dimensional tensors, which preserves diffeomorphism invariance without introducing new degrees of freedom
\begin{align}
 S = \frac{1}{16\pi G} \int d^4x \sqrt{-g} & \Big( \nabla_\mu \nabla_\nu \phi \nabla^\mu \nabla^\nu \phi - c_1 (\Box\phi)^2 + c_2 \tilde{R} + c_3 \tilde{R}^2 + c_4 \tilde{R}_{\mu\nu}\tilde{R}^{\mu\nu} \nonumber \\
 & + c_5 \tilde{C}^\mu{}_{\nu} \tilde{C}_\mu{}^{\nu} + \dots + \lambda (g^{\mu\nu}\partial_\mu\phi\partial_\nu\phi - 1) \Big), 
\end{align}
where $\eta^{\mu\nu\rho\sigma} = \frac{1}{\sqrt{-g}}\epsilon^{\mu\nu\rho\sigma}$. In the projectable Horava model, the lapse function $N$ is only a function of time, $N = N(t)$, and it coincides with the aforementioned family of actions when expressed in the synchronous gauge, $ds^2 = N^2 dt^2 - \gamma_{ij} \left(dx^i + N^i dt\right) \left(dx^j + N^j dt\right)$. The projectable Horava models were shown to be renormalizable \cite{Barvinsky1, Barvinsky2}, and we expect the above Horava-mimetic action to be renormalizable. \\

\noindent In \cite{malaeb2023mimetic}, it was shown that the gravitational action, which includes the necessary Gibbons-Hawking-York (GHY) boundary term, is equivalent to the mimetic Ho\v{r}ava action for spacetimes that can be globally separated into time and space, specifically those with a topology of $R \times \Sigma$. In all other spacetimes, a key difference emerges: the mimetic Ho\v{r}ava action remains well-defined without requiring any surface terms. The stability of cosmological perturbations with a mimetic Ho\v{r}ava gravity model was investigated in \cite{russ2021stability}. 

\section{Asymptotically Free Mimetic Gravity}
\label{section10}

The concept of limiting curvature, previously discussed in Section \ref{section6}, serves as an effective gravitational field theory aimed at preventing spacetime singularities within a primarily classical context. Various models (see \cite{asymptotic1, asymptotic2, asymptotic3}) exhibit asymptotic freedom, a state where the interaction between gravity and matter disappears at the curvature limit. To incorporate limiting curvature in a generally covariant theory based strictly on the metric, one must usually impose constraints on higher-order curvature invariants. This is frequently achieved by including action terms that are quadratic in the metric's second derivatives.\\

This section explores a model of classical asymptotic freedom within the context of mimetic gravity \cite{asymptoticallyfree}. This is implemented by assuming that the gravitational constant, $G = G(\Box \phi)$, is a function of $\Box \phi$ and vanishes at some limiting curvature. A primary difference between this and other frameworks for asymptotic freedom lies in the choice of the scale-dependent variable. For instance, while the model in \cite{asymptotic1} relates the varying gravitational coupling to matter energy density, in the theory presented here, it depends on a particular measure of curvature. This particular construction is carefully designed to modify Einstein's equations without invoking higher-order time derivatives, thereby bypassing the unphysical ghost instabilities often associated with such terms.\\

Within this framework, singularities in contracting flat Friedmann and Kasner universes are resolved \cite{asymptoticallyfree}, with the spacetime transitioning to a de Sitter solution near the limiting curvature. Notably, it also carries the consequence that quantum metric fluctuations asymptotically vanish at this limit, effectively restoring a classical spacetime description. This presents a potential pathway to resolve general relativity's singularity problem through a classical modification at high curvatures, potentially lessening the immediate need for a full non-perturbative quantum gravity theory for such issues. In \cite{asymptoticallyfree}, the Lagrangian took the form
\begin{equation}
\mathcal{L} = f[\phi] R[g_{\mu\nu}] + 2\Lambda[\phi],
\label{lagdenphi}
\end{equation}
where $f$ and $\Lambda$ can depend on $\phi$ and its derivatives, representing the inverse gravitational constant and cosmological constant. To establish the possible dependence of these functions on the scalar field $\phi$ and its derivatives, we must first identify the covariant quantities that can be constructed from $\phi$ \cite{nonflatuniverse}. Due to the constraint
equation \eqref{constraint}, $\phi$ itself can be used as the time coordinate $t$ in a synchronous reference frame, characterized by the metric \cite{landau}
\begin{equation*}
ds^2 = dt^2 - \gamma_{ab}\, dx^a dx^b.
\end{equation*}
Consequently, if $f$ and $\Lambda$ are defined exclusively as functions of $\phi$, the model effectively describes a time-dependent background.
Furthermore, the constraint implies that any covariant quantity derived solely from the first derivatives of $\phi$ is constant; thus, such terms are insufficient for generating a dynamical $f$ and $\Lambda$. In contrast, the second covariant derivatives of $\phi$ are far more significant, as they provide a measure of curvature related to the conformal degree of freedom within the gravitational sector. Specifically, this relationship is given by
\begin{equation*}
-\phi_{;ab} = \kappa_{ab} = \frac{1}{2} \frac{\partial}{\partial t}\gamma_{ab},
\end{equation*}
where $\kappa_{ab}$ is the extrinsic curvature of the slices of constant $\phi$, and $\phi_{;0\alpha}$ is zero. In this synchronous slicing, the Ricci scalar is 
\begin{equation*}
-R = 2\dot{\kappa} + \kappa^2 + \kappa^a_b \kappa^b_a + {}^{(3)}\!R, 
\end{equation*}
where dot represents time derivatives, $\kappa^a_b = \gamma^{ac}\kappa_{cb}$, ${}^{(3)}\!R$ is the 3-curvature of the spatial slices, and
\begin{equation*}
\kappa := \gamma^{ab}\kappa_{ab} = g^{\alpha\beta}\phi_{;\alpha\beta} = \Box\phi 
\end{equation*}
is the trace of extrinsic curvature. This implies that the only way to make the gravitational constant dependent on curvature is by taking $f[\phi] = f(\Box\phi)$. Therefore, equation \eqref{lagdenphi} now takes the form
\begin{equation}
\mathcal{L} = f[\Box \phi] R + 2\Lambda[\Box \phi].
\end{equation}
This theory is free from higher time derivatives in the synchronous frame; however, it was found to still include higher-order mixed and spatial derivatives in the general spatially non-flat case \cite{Zheng, zheng2021hamiltonian}. In fact, the higher-order mixed derivatives in the theory $f(\Box \phi)R+2 \Lambda (\Box \phi)$ were considered in \cite{Zheng} in an attempt to correct of the sign of the mimetic scalar field's gradient term in a flat Friedmann universe. However, later, in \cite{zheng2021hamiltonian}, it was shown that the theory contains an additional, hidden scalar degree of freedom that introduces new instabilities once perturbations are introduced.\\

\noindent The action, free of higher derivatives, was first introduced in \cite{remnants}. The term responsible for higher mixed and spatial derivatives can be covariantly removed (see \cite{nonflatuniverse} for details), resulting in the action
\begin{equation}
    S = \frac{1}{16 \pi} \int d^4 x \sqrt{-g} \left( - \mathcal{L} + \lambda \left(g^{\mu \nu} \partial_{\mu} \phi \partial_{\nu} \phi - 1 \right) \right),
\label{nonflataction}
\end{equation}
where 
\begin{equation*}
    \mathcal{L} = f(\Box \phi) R + (f(\Box \phi) - 1) \tilde{R} + 2 \Lambda(\Box \phi),
\end{equation*}
and $\tilde{R}$ coincides with $ \prescript{(3)}{} R$ in synchronous gauge
\begin{align*}
    \tilde{R} = 2 R^{\mu\nu} \partial_{\mu} \phi \partial_{\nu} \phi - R - (\Box \phi)^2 + \nabla_{\mu} \nabla_{\nu} \phi \nabla^{\mu} \nabla^{\nu} \phi.
\end{align*}
This theory allowed for the finding of an exact black hole solution. Furthermore, in \cite{nonflatuniverse}, the singularity resolution in cosmology and black hole physics were investigated, where it was shown that the Big Bang singularity can be replaced with a regular bounce in homogeneous and isotropic universes (flat, open, and closed), and the central singularity of Schwarzschild-like black holes can be resolved. In the latter, to enhance generality, a nonlinear spatial curvature-dependent potential $h(\tilde{R})$ was added to the Lagrangian density
\begin{equation}
    \mathcal{L} = f(\Box \phi) R + (f(\Box \phi) - 1) \tilde{R} + 2 \Lambda(\Box \phi) + h(\tilde{R}).
\label{lagden}
\end{equation} 
This addition reintroduces higher spatial derivatives into the theory, but crucially, it avoids higher time or mixed derivatives. This is a significant distinction because higher time derivatives often lead to problematic extra degrees of freedom, such as ghosts, which can make a theory unstable. In contrast, these higher spatial derivatives could be beneficial, as they have the potential to improve the renormalizability of gravity, similar to the approach taken in Ho\v{r}ava gravity \cite{mimetichorava}.\\

\noindent Upon varying the action (equation \ref{nonflataction}) with $\mathcal{L}$ given by \eqref{lagden}, the modified Einstein equations are given by (for details, check \cite{nonflatuniverse} and appendix B therein)
\begin{equation*}
\text{mixed components}: \quad f R_{0a} + Z_{,a} + \kappa^b_a f_{,b} = 8\pi T^{(m)}_{0a},
\end{equation*}
\begin{equation*}
\text{spatial components}: \quad- \frac{1}{\sqrt{\gamma}} \partial_t \left( \sqrt{\gamma} \left( f \kappa^a_b + Z \delta^a_b \right) \right) - \frac{1}{2} \mathcal{L} \delta^a_b = S^a_b + 8\pi T^{(m)a}_{\quad b},
\end{equation*}
\begin{equation}
\text{time-time components}: \quad \frac{1}{3} (f - 2\kappa f') \kappa^2 - \Lambda + \kappa \Lambda' - \frac{1}{2} (f + \kappa f') \tilde{\kappa}^a_b \tilde{\kappa}^b_a = \frac{1}{2} (h - {}^{(3)}\!R) + \Xi + 8\pi T^{(m)}_{00},
\label{temporal}
\end{equation}
where
\begin{align*}
{}& S^a_b = (1-h^I) {}^{(3)}\!R^a_b + h^{I a}_{\;\; ;b} - \Delta h^I \delta^a_b \\
& \Xi = \frac{1}{\sqrt{\gamma}} \int dt \sqrt{\gamma} \left( T^{(m)a}_{\quad 0} - (1-h')R^a{}_0 + \kappa h^{',a} - \kappa^a_b h^{',b} \right)_{|a}
\end{align*}
and
\begin{equation*}
f' := df/d\Box\phi, \quad \Lambda' := d\Lambda/d\Box\phi, \quad h^I := dh/d\tilde{R} \quad \text{and} \quad T^{(m)}_{\mu\nu} = \frac{2}{\sqrt{-g}} \frac{\delta S^m}{\delta g^{\mu\nu}}.
\end{equation*}

\subsection{Non-flat Universes}

In a flat, isotropic universe, asymptotic freedom is a choice, while for resolving anisotropic Kasner singularities, it is a necessity driven by how the function $f$ affects the evolution of anisotropies. In standard GR, anisotropies scale as $1/a^6$ near a singularity, while in this theory, this scaling is altered by a running gravitational coupling $G(\kappa)$. By analyzing the Bianchi type I (anisotropic) universe, which includes the Kasner solution, it was shown in \cite{limitingcurvature} that for $\kappa^2$ to remain bounded and avoid a singularity as the universe contracts ($a \to 0$), the functions $f$ and $G$ must exhibit asymptotic freedom. \\

\noindent Considering first the Friedmann universe, which describes a homogeneous and isotropic universe with cosmological time t 
\begin{equation*}
    ds^2 = dt^2 - a^2(t) \left( \frac{dr^2}{1 - \mathcal{\chi} r^2} + r^2 \left( d \theta^2 + sin^2 \theta \ d\phi^2 \right) \right),
\end{equation*}
where $\mathcal{\chi} \in \{-1, 0 , 1\}$. For the spatially flat universe, $\chi = 0$, instead of Big Bang/Big Crunch singularities, the non-singular modification introduces a smooth transition, leading to (or emerging from) an initial/final de Sitter-like state characterized by limiting curvature. For a spatially non-flat case, $\chi = \pm 1$, one can show that a contracting universe will first experience a phase of exponential contraction, then pass through a bounce, and subsequently enter a period of exponential expansion. The duration of this inflationary stage can be estimated, and a necessary number of e-folds can be obtained. In specific, for a closed universe ($\chi = 1$) and a choice for the function $f$
\begin{equation*}
f = \frac{1}{1 - \left( \frac{\kappa^2}{\kappa_0^2} \right)},
\end{equation*}
where $\kappa = 3 \frac{\dot{a}}{a}$ and $\kappa_0$ is some limiting value, we get a different solution from that of standard general relativity. In standard GR, we get a closed universe originating from a Big Bang, expanding, then recollapsing until finally reaching a Big Crunch. In this theory, where $\kappa$ becomes of order of the limiting curvature at $t - a_{\text{max}} \sim 1/\kappa_0$ \footnote{$a_{max}$ represents the maximum value the scale factor $a(t)$ attains and it corresponds to the point where the universe stops expanding and begins to contract (moment of recollapse).}, modifications starts to take over and we get a smooth transition. Instead of a Big Crunch, the scale factor, after reaching $a_{\text{max}}$, starts to decrease exponentially due to curvature effects. Crucially, the spatial curvature term eventually counteracts this exponential contraction, preventing unbounded growth of spatial curvature. A smooth bounce occurs at $t = t_b$ (moment of bounce) where $a=a_{\text{min}}$ \footnote{This represents the minimum value the scale factor $a(t)$ reaches and it is the point of the non-singular bounce.}. We transition from the contracting half-plane to the expanding half-plane. Post-bounce, the universe expands, mirroring the contracting phase, and grows until $\kappa$ reaches $\kappa_0$, undergoes a period of graceful exit, and eventually, the spatial curvature term dominates again, leading to another recollapse. This results in an eternally oscillating, non-singular universe, following a closed trajectory in phase space.\\

This model was also extended to various Bianchi universes,  which are spatially homogeneous but generally anisotropic, to investigate singularity resolution \cite{nonflatuniverse}. For Bianchi Type I (spatially flat, anisotropic), resolving the Kasner singularity is shown to necessitate asymptotic freedom, where the modifying function $f$ should diverge fast enough at the limiting curvature. Bianchi Type V (spatially non-flat with isotropic spatial curvature) exhibits similar conditions for singularity resolution. More complex types (II, VI$_0$, VII$_0$, VIII, IX) are treated collectively; depending on a spatial curvature-related constant $d$, these models can lead to either a single bounce with no recollapse (if $d>0$, like Type V or VII$_0$) or, crucially, to non-singular, eternally oscillating cyclic universes (if $d<0$, like Type IX), where both curvature and anisotropic scale factors oscillate between bounds. The evolution of a non-flat universe was also explored in \cite{casalino2020higher} under the context of higher-derivative and mimetic modified gravity models.

\subsection{Black Hole Remnants}

This subsection discusses the proposition in \cite{remnants, nonflatuniverse} to the resolution of the black hole singularity problem and black hole evaporation within a classical modified gravity framework incorporating a limiting curvature principle. We will see that the singularity will be replaced by a transition to a static de Sitter patch at limiting curvature. What is remarkable is that the transition happens smoothly and gradually, which makes it distinct from most of the existing non-singular black modifications \cite{non-singblack2, non-singblack3, non-singblack4, asymptotic3}. \\

\noindent We start from the metric for both the black hole and the de Sitter universe in static coordinates, 
\begin{equation}
ds^2 = (1 - a^2(r)) dt^2 - \frac{dr^2}{(1 - a^2(r))} - r^2 \left( d\vartheta^2 + \sin^2 \vartheta d\varphi^2 \right),
\label{bh&dS}
\end{equation}
where for a black hole of mass $M$, the function $a^2(r)$ is defined as $a^2(r) = r_g/r$, where the Schwarzschild radius is $r_g = 2M$. In the case of a de Sitter universe, this function becomes $a^2 = (Hr)^2$, with $H^{-1}$ representing the radius of curvature. We see that at the horizon $(a = 1)$, there is a coordinate singularity. Using instead the synchronous Lema$\hat{i}$tre coordinate  \cite{lemaitre}, which are non-singular on the horizons, 
\begin{align*}
T &= t + \int \frac{a}{1 - a^2} \, dr, \quad 
R = t + \int \frac{dr}{a(1 - a^2)},
\end{align*}
the metric becomes in the new coordinates
\begin{align}
ds^2 = dT^2 - a^2(x)\,dR^2 - b^2(x)\,d\Omega^2.
\label{metric}
\end{align}
Using the relation 
\begin{align}
x \equiv R - T = \int \frac{dr}{a(r)},
\label{xeq}
\end{align} 
$a^2$ and $b^2 = r^2$ must be expressed in terms of $R$ and $T$. The (non-rotating) black hole metric in these coordinates is given by 
\begin{align}
ds^2 = dT^2 - \left( \frac{x}{x_+} \right)^{-2/3} dR^2 - \left( \frac{x}{x_+} \right)^{4/3} r_g^2\, d\Omega^2.
\label{bh}
\end{align}
This metric is regular at the horizon, located at $x = x_+ = 4M/3$. The interior and exterior regions of the black hole are both covered by the domain $x > 0$, while $x = 0$ signifies the physical, space-like singularity where curvature invariants diverge. Consequently, within the framework of general relativity, this metric cannot be extended to negative values of $x$.\\

\noindent The de Sitter metric can be expressed as
\begin{equation}
ds^2 = dT^2 - \exp(2H(x - x_-)) (dR^2 + H^{-2} d\Omega^2),
\label{dS}
\end{equation}
where $x_-$ is a constant of integration arising from equation \eqref{xeq}, and the de Sitter horizon is situated at $x = x_-$. Both solutions presented in equations \eqref{bh} and \eqref{dS} are spatially flat when considered within the Lema$\hat{i}$tre slicing. This can be seen upon calculating the spatial curvature of the metric \eqref{metric}. It is found that for constant $T$ hypersurfaces, the spatial curvature vanishes if $a = \frac{d b}{dx}$.\\

\noindent To see how we can get a non-singular black hole in an asymptotically free mimetic gravity, we consider the metric ansatz \eqref{metric}. Given that the Lema\^{\i}tre coordinates establish a synchronous frame, the constraint equation $\partial_{\mu} \phi_{\alpha} \partial^{\mu} \phi^{\alpha} = 1$ for the mimetic field admits a general solution, $\phi = T + \text{const.}$, where $T$ is the Lema\^{\i}tre time coordinate. This solution respects the symmetries inherent in the metric ansatz \eqref{metric}. Therefore, using the action given in \eqref{nonflataction}, we can get the modified Einstein equations for the metric \eqref{metric}. Those equations of motion can be solved to get the functions $a$ and $b$.\\

\noindent For a contracting Kasner universe, we saw in the previous subsection that the vanishing gravitational constant can make anisotropies disappear during contraction, and the solution transitions towards a de Sitter universe near the limiting curvature. We expect here the black hole singularity to be resolved in a similar manner. In the standard Schwarzschild solution \eqref{bh}, the function $a$ follows $a \propto b^{-1/2}$ and diverges as $b \to 0$, whereas it behaves as $a \propto b$ in the de Sitter-like solution. This implies that $a$ must reach a maximum value at some point $x_*$ inside the black hole before decreasing again as we deeper into the black hole. If this maximum value of $a$, $a(x_*)$, is greater than one, two Killing horizons ($x_\pm$) exist where $a(x_\pm)=1$ at $x_+ > x_*$ and $x_- < x_*$ in analogy with equations \eqref{bh} and \eqref{dS}. As the black hole mass decreases, these horizons approach each other and merge when the maximum of $a$ is exactly one, corresponding to a minimal black hole mass $M_{\text{min}} \sim 1/\kappa_0$. Below this $M_{\text{min}}$, no horizon forms, and thus no black hole solution exists. Crucially, the surface gravity $g_s = -a'(x_\pm)$ (where prime denotes the derivative with respect to x) vanishes for this minimal mass black hole. Since Hawking radiation temperature is proportional to $g_s$, it also becomes zero for these minimal mass remnants. This indicates that black hole evaporation will cease when the black hole reaches this minimal mass $M_{\text{min}}$, leaving behind a stable remnant. Therefore, the combination of limiting curvature and the principle of asymptotic freedom in their gravitational theory generically predicts the existence of such minimal, stable black hole remnants.\\

\noindent An exact solution can be obtained (even for $h \neq 0$) within the spatial flatness approximation; this solution strictly adheres to the spatial flatness condition $\left( a = \frac{db}{dx} \right)$ and describes a non-singular black hole with a stable remnant. The function $f(\chi)$ is chosen to be asymptotically free
\begin{align*}
    {}& f(\kappa) = \frac{1+3\left(\frac{\kappa}{\kappa_0}\right)^2}{ \left(1 + \left( \frac{\kappa}{\kappa_s} \right)^2 \right) \left(1 - \left( \frac{\kappa}{\kappa_0} \right)^2 \right)^2},
\end{align*}
where $\lim_{\kappa \to \infty} f(\kappa) \rightarrow 0$, and $\Lambda(\kappa)$ is chosen to be
\begin{align*}
\Lambda(\kappa) {}&= \kappa^2 \left( \frac{\frac{4}{3} \left(\kappa/\kappa_0\right)^2}{\left(1 - (\kappa/\kappa_0)^2\right)^2} - \frac{1 + 2 (\kappa/\kappa_0)^2}{1 + 4 (\kappa/\kappa_0)^2 + 3 (\kappa/\kappa_0)^4} \right) \\
& + \frac{\kappa_0}{6} \kappa \left( \arctan \frac{\kappa}{\kappa_0} - 3 \sqrt{3} \arctan \left( \sqrt{3} \frac{\kappa}{\kappa_0} \right) + 2 \, \text{atanh} \frac{\kappa}{\kappa_0} \right).
\end{align*}
After some algebra, the equations of motion can be solved to get the solutions
\begin{align*}
    {}& a^3 (\kappa) = \frac{4M}{3} \mid \kappa\mid \left( 1 - \left(\frac{\kappa}{\kappa_0} \right)^4 \right) \left( \frac{1+\left( \frac{\kappa}{\kappa_0} \right)^2}{1+ 3\left( \frac{\kappa}{\kappa_0} \right)^2 } \right)^2, \\
    & b^3 (\kappa) = \frac{9M}{2 \kappa^2} \left( 1 - \left(\frac{\kappa}{\kappa_0} \right)^2 \right) \left( 1 + 3 \left(\frac{\kappa}{\kappa_0} \right)^2 \right), \\
    & H(\kappa) = \ln \left( -\frac{\kappa}{\kappa_0} \frac{1+\left( \frac{\kappa}{\kappa_0} \right)^2}{1+ 3\left( \frac{\kappa}{\kappa_0} \right)^2 } \right).
\end{align*}
Using the temporal modified Einstein equation \eqref{temporal}, we obtain a first order differential equation that can be solved to express $\kappa$ in term of $x$. From the latter, we can see that in the far exterior region where $x \to \infty$, we have $ \left( \frac{\kappa}{\kappa_0} \right)^2 \to 0$, and the solution approaches a black hole with mass $M$. On the other hand, deep inside the black hole, as $x \to -\infty$, we get $\tilde{\kappa}^2 \to 1$, and the solution corresponds to a de Sitter space with $H = \kappa_0 / 3$. This means that the exact solution smoothly connects the two limits. For this solution, $a$ is maximized at $ \kappa = \kappa_* = -\kappa_0 / \sqrt{5} $. At $\kappa_*$, the value of $a$ is given by 
\begin{equation*}
a(\kappa_*) = \left( \frac{18 \kappa_0}{25 \sqrt{5}} M \right)^{1/3} =: \left( \frac{M}{M_{\text{min}}} \right)^{1/3},
\end{equation*}
and the minimal possible black hole mass in this case is
\begin{equation}
M_{\text{min}} = \frac{25 \sqrt{5}}{18 \kappa_0}.
\end{equation}
Black holes have a lower mass limit, and at this minimum, they possess a single horizon with zero Hawking temperature. Therefore, black hole evaporation ultimately leaves behind a stable remnant with minimal mass. By going back to the coordinates given in \eqref{bh&dS}, we see that at large distances from the black holes ($r \to \infty$), the solution closely approximates the standard Schwarzschild metric and the curvature becomes comparable to the limiting curvature at $r_* = \left(144 M / 5 \kappa_0^2 \right)^{(1/3)}$. For massive black holes with $M \gg M_{\text{min}}$, there exists an outer horizon at $r_+$
\begin{equation*}
r_+ = 2M \left[ 1 - \mathcal{O}\left(\left(\frac{M_{\text{min}}}{M}\right)^2\right) \right],
\end{equation*}
and deviations from the Schwarzchild solution are significant only deep inside the black hole where $r \ll r_+$. Near the limiting curvature, the metric is well-approximated by a de Sitter region, which features an inner horizon at $r_-$
\begin{equation}
r_- = H^{-1} \left[ 1 + \mathcal{O}\left(\frac{M_{\text{min}}}{M}\right) \right].
\end{equation}
Nevertheless, these asymptotic approximations fail when the black hole's mass $M$ is comparable to the minimal mass $M_{\min}$. In the specific case where $M = M_{\min}$, the inner and outer horizons coincide to form a single horizon. This resulting geometry is reminiscent of the near-horizon metric of an extremal charged Reissner--Nordstr\"{o}m black hole; however, these solutions remain non-singular and are inherently stable.\\

\noindent In \cite{nonflatuniverse}, the work was extended for more general black hole solutions where small amounts of electric charge or angular momentum were included. The findings were that in this framework, the black holes are similar to Reissner-Nordström or Kerr black holes but without their problematic instabilities. There is an inner horizon and a static de Sitter core instead of a singularity. 

\section{Mimetic Inflation}
\label{section11}

The mimetic field framework provides a unique resolution to several long-standing issues in inflationary cosmology. Specifically, one can construct inflationary models where the problematic regime of self-reproduction is avoided, even when inflation begins at the Planck scale. This section explores two major developments in this area: the use of potentials coupled to curvature functions to suppress the formation of an eternal multiverse \cite{mimeticinflation}, and the coupling of the mimetic field to topological invariants to resolve the exponential dilution of dark matter \cite{chamseddine2026mimetic}.

\subsection{Inflation and Self-Reproduction}

In standard slow-roll inflation, quantum fluctuations at the Hubble scale can exceed the classical descent of the inflaton field, leading to an eternally inflating multiverse and a loss of predictability. In \cite{mimeticinflation}, it was shown that this can be avoided by coupling the inflation potential to the mimetic field $\phi$. This does not imply the presence of an additional scalar field, but rather represents a modified Einstein gravity at high curvatures. The proposed action is 
\begin{equation*}
    S = \int d^4 x \sqrt{-g} \left(-\frac{1}{2} R + \lambda \left( g^{\mu \nu} \partial_{\mu} \phi \partial_{\nu} \phi - 1 \right) + \frac{1}{2}g^{\mu \nu} \partial_{\mu} \varphi \partial_{\nu} \varphi  - C \left(\kappa\right) V\left(\varphi\right)\right),
\end{equation*}
where $\kappa \equiv \Box \phi = g^{\mu \nu}\phi_{; \mu \nu}$ in the synchronous coordinate system and $\varphi$ is the inflaton field with potential $V(\varphi)$ coupled to a function $C(\kappa)$. This is similar to Higgs coupling to curvature $R$, $\zeta \bar{H} H R$ when $H$ is used an inflaton. However, notice the simplicity of interactions of inflaton to extrinsic curvature (linear), not possible in GR where $R$ depends on $\kappa^2$, $\dot{\kappa}$, and $\kappa_i^j \kappa_j^i$, forcing a rescaling of the metric to go to the Einstein frame. Upon varying the action with respect to the Lagrange multiplier, we get the constraint equation, while the $\phi$ equation is given by 
\begin{equation*}
    \partial_\mu \left[ \sqrt{-g} g^{\mu\nu} \left( 2\lambda \partial_\nu \phi + \partial_\nu (C' V) \right) \right] = 0,
\end{equation*}
where $C' \equiv dC/d\kappa$. The inflation equation is obtained by varying with respect to $\varphi$ and is given by 
\begin{equation}
    \frac{1}{\sqrt{-g}} \partial_\mu \left( \sqrt{-g} g^{\mu\nu} \partial_\nu \varphi \right) + C V' = 0,
\label{inflatoneq}
\end{equation}
where the first term is nothing but $\Box \varphi$ and $V' \equiv dV/d \varphi$. Varying the action with respect to the metric, we get the modified Einstein equations
\begin{equation*}
G_{\mu\nu} = R_{\mu\nu} - \frac{1}{2} g_{\mu\nu} R = T_{\mu\nu},
\end{equation*}
where
\begin{align*}
    T_{\mu\nu} &= 2\lambda \partial_\mu \phi \partial_\nu \phi + g_{\mu\nu} \left( (C - \kappa C')V - g^{\rho\sigma} \partial_\rho (C' V) \partial_\sigma \phi \right) \nonumber \\
    &\quad + \left( \partial_\mu (C' V) \partial_\nu \phi + \partial_\nu (C' V) \partial_\mu \phi \right) + \partial_\mu \varphi \partial_\nu \varphi - \frac{1}{2} g_{\mu\nu} \left( g^{\rho\sigma} \partial_\rho \varphi \partial_\sigma \varphi \right).
\end{align*}
Considering a homogeneous flat universe with metric
\begin{equation*}
ds^2 = dt^2 - a^2(t)\, \delta_{ik} dx^i dx^k,
\end{equation*}
the equations of motion can be solved, and they simplify to 
\begin{equation}
    \frac{1}{3}\kappa^2 = (C - \kappa C') V + \frac{1}{2}\dot{\varphi}^2,
\label{eq1}
\end{equation}
and
\begin{equation}
    \ddot{\varphi} + \kappa\dot{\varphi} + CV' = 0,
\label{eq2}
\end{equation}
where $\kappa$ is the tripled Hubble constant, $\kappa = 3 \dot{a}/a$.\\

\noindent For a model consistent with CMB observations, the potential tends to be flat at observable scales, and the tensor-to-scalar ratio must be small. There are many potentials that satisfy these requirements, and they do not exhibit self-reproduction. Namely, we choose a simple model with the function corresponding to the mimetic interactions and the potential given respectively by
\begin{align}
    {}& C(\kappa) = 1 + \frac{\kappa}{m} \nonumber \\ 
    & V(\varphi) = \frac{1}{2} \frac{m^2 \varphi^2}{1+\varphi^2} \left(1+m \varphi^4 \right), \label{modeleq}
\end{align}
and we assume $m \ll 1$ in Planck units. For small field values, $\varphi < 1$, this potential describes a massive scalar field. Conversely, for large field values, $\varphi > 1$, it can be well approximated by the expression
\begin{equation}
V \approx \frac{1}{2}m^2 \left( 1 - \frac{1}{\varphi^2} + m\varphi^4 \right). 
\label{eq:potential_approx}
\end{equation}
Under this approximation, the system of equations \eqref{eq1} and \eqref{eq2} simplifies to:
\begin{equation*}
\kappa^2 = \frac{3}{2}m^2 \left( 1 - \frac{1}{\varphi^2} + m\varphi^4 \right) + \frac{3}{2}\dot{\varphi}^2,
\end{equation*}
and
\begin{equation*}
\ddot{\varphi} + \kappa\dot{\varphi} + \left( 1 + \frac{\kappa}{m} \right) \left( \frac{m^2}{\varphi^3} + 2m^3\varphi^3 \right) = 0. 
\end{equation*}
By analyzing these equations and comparing the potential and kinetic terms, we find that up to the Planck scale—reached at $\varphi \approx m^{-2/3}$, self-reproduction does not occur. Therefore, we can begin inflation at the Planck scale, simultaneously resolving the fine-tuning problem and preventing an eternally self-reproducing universe. \\

\noindent One should also analyze the cosmological perturbations in their mimetic inflation model. We start by writing the perturbed metric in the conformal Newtonian gauge \cite{newtoniangauge}
\begin{equation*}
    ds^2 = (1 + 2\Phi)\,dt^2 - a^2(t) \left[ (1 - 2\Phi)\,\delta_{ik}\,dx^i dx^k - h^{(t)}_{ik}\,dx^i dx^k \right],
\end{equation*}
where $\Phi$ is the gravitational potential and $h_{ik}^{(t)}$, the transverse traceless part of the metric, represents gravitational waves. Although the equation governing gravitational waves remains identical to that in general relativity, the consideration of scalar perturbations in our mimetic inflation model undergoes significant modifications. The constraint equation of the mimetic field $\phi$ simplifies to
\begin{equation}
    \delta\phi = \Phi    
\label{contraintperturbation}
\end{equation}
at the linear level, directly linking the field perturbations to the gravitational potential. The $0 - i$ Einstein field equation will give
\begin{align}
    \dot{\Phi} + H \Phi &= \frac{1}{2\left(1 + \frac{3}{2} C'' V\right)} 
    \left[ \left( \dot{\varphi} + C' V' \right) \delta\varphi 
    - \frac{C'' V}{a^2} \Delta \delta\phi \right] \nonumber \\
    &= -\frac{\dot{H}}{\dot{\varphi}} \delta\varphi 
    - \frac{C'' V}{\left(2 + 3 C'' V\right) a^2} \Delta \delta\phi.
\label{0-iperturbation}
\end{align}
By linearizing the inflaton equation (eq. \ref{inflatoneq}), it simplifies to 
\begin{equation}
\ddot{\delta\varphi} + 3H\dot{\delta\varphi} - \frac{1}{a^2} \Delta \left( \delta\varphi + \frac{2C'V' - 4\dot{\varphi}C''V}{2 + 3C''V} \delta\varphi \right) 
- \left( \frac{\ddot{\varphi}}{\dot{\varphi}} + 3H\frac{\dot{\varphi}}{\dot{\varphi}} - \dot{H} \right) \delta\varphi 
- 2(\ddot{\varphi} + H\dot{\varphi}) \Phi = 0.
\label{inflatonpertubation}
\end{equation}
Therefore, we obtain a system of coupled differential equations for the perturbations (equations \eqref{contraintperturbation}, \eqref{0-iperturbation}, and \eqref{inflatonpertubation}). These three equations are enough to fully determine the unknown functions $\delta\varphi, \delta\phi, \Phi$. The other field equations, $0-0$ and $i-i$, do not provide any additional useful information. Consider the plane wave with co-moving wave-number $ k = \left| \vec{k} \right| $, i.e., 
\begin{equation*}
   \delta\varphi, \Phi, \delta\phi \propto \exp\left( i \vec{k} \cdot \vec{x} \right).
\end{equation*}
After analyzing the behavior of the perturbations, which depends drastically on whether the physical wavelength \( \lambda_{\text{ph}} \simeq a/k \) is much smaller or much larger compared to the Hubble scale \( H^{-1} \), we see that for short-wavelength perturbations ($k \gg Ha$), the system simplifies, and the solution for $\delta\phi_k$ is found to be a decaying mode
\begin{equation*}
    \delta\varphi_k \simeq \frac{A_k}{a} \exp\left( \pm i k \int \frac{dt}{a} \right).
\end{equation*}
However, for long-wavelength perturbations ($k \ll Ha$), which are relevant for the formation of large-scale structure, the spatial derivative terms decay and can be neglected. Next, the spectrum of inhomogeneities generated by initial quantum fluctuations during mimetic inflation can be computed. It was found that there is no fine-tuning problem. Any initial particles or inhomogeneities exist on very small scales. As long as these initial imperfections aren't large enough to stop the inflationary expansion from starting, the expansion itself will stretch them to unobservably large scales, making them completely irrelevant. Therefore, inflation can successfully start even within a highly inhomogeneous region, provided that the initial inhomogeneities are not strong enough to obstruct the onset of inflationary expansion. One can also see that a fine-tuned initial vacuum state, such as the Bunch-Davies vacuum, is not required. Instead, the only necessary condition is that any initial inhomogeneities on sub-Hubble scales do not prevent the onset of inflation. The model presented above (equations \eqref{modeleq}) is consistent with cosmological data and does not lead to problems related to an eternal universe. By considering $1 < \phi < m^{-1/6}$ and introducing the number of e-folds $N_k$ before the end of inflation
\begin{equation*}
    a \simeq a_f e^{-N},
\end{equation*}
where $a_f$ is the scale factor at the end of inflation, the model yields a scalar spectral index $n_s - 1 = -3/(2N_k)$ and a tensor-to-scalar ratio $r \propto 1/N_k^{3/2}$. For the observationally relevant range $50 < N_k < 60$, this correctly produces $n_s \approx 0.97$ (for $N_k = 50$) in agreement with current observational constraints. This model also ensures that perturbation amplitudes remain less than unity below the Plank scale, avoiding the issue of self-reproduction.\\

\noindent We end this section by emphasizing that no additional scalar fields were added; instead, it represents a minimal modification of Einstein gravity. The mimetic field introduces only one extra degree of freedom, a “dust” component, which rapidly becomes negligible after inflation but can reappear later as a candidate for dark matter. However, a significant theoretical limitation of this framework is the arbitrary nature of the coupling between the mimetic field and the inflaton potential. This specific construction is somehow ad hoc.

\subsection{Mimetic Dark Matter from Gauss-Bonnet Inflation}

A persistent challenge in mimetic cosmology is the exponential dilution of dark matter during inflation. If dark matter is merely an integration constant, its density scales as $a^{-3}$, becoming negligible after $60$ e-folds of expansion. To provide a natural, geometric solution, Chamseddine and Mukhanov \cite{chamseddine2026mimetic} proposed coupling the mimetic field to the Gauss-Bonnet (GB) term 
\begin{align*}
    \mathcal{G} = R_{\mu\nu\alpha\beta}R^{\alpha\beta\mu\nu} - 4R_{\mu\nu}R^{\mu\nu} + R^2.
\end{align*}
The action takes the form
\begin{equation*}
S = \int d^4x \sqrt{-g} \left[ -\frac{1}{2}R + \lambda(g^{\mu\nu}\partial_\mu\phi\partial_\nu\phi - 1) + f(\phi)g(\Box\phi) \mathcal{G} \right].
\end{equation*}
In four dimensions, $\mathcal{G}$ is a total derivative, which can be written as a Chern-Simons three-form. While it does not contribute to the equations of motion in standard General Relativity, its coupling to functions of the mimetic field $f(\phi)$ and its d'Alembertian $g(\Box\phi)$ allows it to source the field equations dynamically without introducing ghosts.\\

\noindent In a Friedmann universe, the $0-0$ Einstein equation is modified by the GB contribution
\begin{equation*}
3H^2 = 2\lambda - 24 \dot{f} g H^3 - 48 f g' \dot{\kappa} H^2 + 24 (\dot{f} g' + f \dot{g}') \dot{\kappa} H + 24 f g' H^2 \left( \dot{H} + 3H^2 \right).
\end{equation*}
Through the variation of the scalar field $\phi$, the Lagrange multiplier $\lambda$ is found to be
\begin{equation*}
2\lambda = \frac{8}{a^3} \int dt f'(\phi) g(\kappa) \partial_0 (a^3 H^3) + 24 \partial_0 (f g' H^2 \dot{\kappa}).
\end{equation*}
For the well-motivated choice $f(\phi) = \beta \phi$ and $g(\kappa) = \kappa^3$, the integral can be solved analytically for the end of the inflationary phase. The resulting Friedmann equation during the subsequent radiation-dominated era is
\begin{equation*}
3H^2 = 162 \beta H_I^6 \frac{a_f^3}{a^3} + 3H_I^2 \frac{a_f^3 a_{rad}}{a^4},
\end{equation*}
where $H_I$ is the Hubble parameter at the end of inflation. This proves that dark matter is not an initial condition but is \textit{seeded} by the expansion itself. By matching observations, the authors find that matter-radiation equality occurs at 
\begin{align*}
    a_{eq}/a_{rad} = (54\beta)^{-1} H_I^{-4}. 
\end{align*}
In Planck units, $H_I \sim 10^{-6}$ and for $\beta \sim 0.1$, the theory yields $t_{eq} \sim 10^{12}$ s, in excellent agreement with the observed universe.\\

\noindent The authors also explore ``Anomalous Dark Matter'' by considering non-linear functions like $f(\phi) \propto \phi^2$. This modifies the expansion rate to $a \propto t^n$ where $n = \frac{2}{3}(1 + \alpha/9)$, providing a potential observational signature for the interaction between dark matter and the inflationary sector. This geometric coupling ensures the theory remains ghost-free and provides a robust unification of the early-universe dynamics and the dark sector.

\section{Conclusion}
\label{section12}

The simple yet profound step of reparametrizing the physical metric $g_{\mu\nu}$ in terms of an auxiliary metric $\tilde{g}_{\mu\nu}$ and a scalar field $\phi$ via the relation
\begin{equation*}
g_{\mu\nu} = \tilde{g}_{\mu\nu} \left( \tilde{g}^{\alpha\beta} \partial_{\alpha} \phi \partial_{\beta} \phi \right),
\end{equation*}
effectively isolates the conformal degree of freedom. By construction, this ensures that the physical metric satisfies the constraint
\begin{equation*}
g^{\mu\nu} \partial_{\mu} \phi \partial_{\nu} \phi = 1,
\end{equation*}
where $\phi$ plays the role of a gravitational potential representing synchronous time. This constrained system provides a remarkably minimal modification of general relativity; while the number of independent degrees of freedom remains essentially the same as in the standard theory, the reparametrization allows for the emergence of ``mimetic matter'' in the form of a pressureless dust. Crucially, this is achieved without the immediate need for higher-order curvature terms like $R^2$, but rather through the internal dynamics of the scalar field.\\

\noindent Throughout this review, we have seen that the mimetic framework is highly versatile. It can be extended to $f(R, \phi)$ theories to unify the dark sector or to $f(\Box \phi)$ models to implement a limiting curvature and resolve cosmological singularities. Additionally, it is worth noting that the mimetic construction is not restricted to purely bosonic actions. It has been shown that the mimetic constraint can be successfully embedded into $\mathcal{N}=1$ supergravity \cite{supergravity}, allowing the mimetic dark matter candidate to be incorporated into a more fundamental high-energy completion without violating the principles of supergeometry. However, the theory is not without its challenges. The presence of gradient instabilities and ghosts in the simplest mimetic models necessitates careful modifications, leading to the development of more robust versions such as mimetic massive gravity and mimetic Ho\v{r}ava gravity.\\

\noindent In summary, mimetic gravity allows us to address fundamental problems in cosmology and high-energy physics-from dark matter and inflation to the resolution of the Big Bang singularity-at a minimal cost to the mathematical structure of the theory. The major challenge remaining for this formulation is to move beyond theoretical elegance and identify a unique, testable observable prediction that can distinguish mimetic gravity from other dark sector and modified gravity models in future high-precision astrophysical and cosmological observations.\\

\bibliography{references}

\end{document}